\documentclass[%
 reprint,
nofootinbib,
 amsmath,amssymb,
 aps,
pra,
]{revtex4-2}

\usepackage{natbib}
\usepackage{comment}
\usepackage[normalem]{ulem} \usepackage{xcolor}
\usepackage{graphicx}
\usepackage{dcolumn}
\usepackage{bm}
\usepackage{hyperref}
\usepackage{dsfont}

\begin{document}

\preprint{APS/123-QED}

\title{Dissipation engineering of fermionic long-range order beyond Lindblad}

\author{Silvia Neri}
\affiliation{Max Planck Institute for Solid State Research, Stuttgart, Germany}

\author{François Damanet}
\affiliation{Institut de Physique Nucléaire, Atomique et de Spectroscopie, CESAM, Universit\'e de Li\`ege, 4000 Liège, Belgium}

\author{Andrew J. Daley}
\affiliation{Department of Physics, University of Oxford, Oxford, UK.}

\author{Maria Luisa Chiofalo}\thanks{Corresponding author: maria.luisa.chiofalo@unipi.it}
\affiliation{Dipartimento di Fisica “Enrico Fermi” and INFN, University of Pisa, Italy}

\author{Jorge Yago Malo}
\affiliation{Dipartimento di Fisica “Enrico Fermi” and INFN, University of Pisa, Pisa, Italy}

\date{\today}

\begin{abstract}
We investigate the possibility of engineering dissipatively long-range order that is robust against heating in strongly interacting fermionic systems, relevant for atoms in cavity QED. It was previously shown [Tindall \textit{et al.} Phys.\ Rev.\ Lett.\ \textbf{123}, 030603 (2019)] that it is possible to stabilize long-range order in a Hubbard model by exploiting a dissipative mechanism in the Lindblad limit, this latter being valid for spectrally unstructured baths. Here, we first show that this mechanism still holds when including additional spin-exchange interactions in the model, that is for the tUJ model. Moreover, by means of a Redfield approach that goes beyond the Lindblad case, we show how the stability of the engineered state depends crucially on properties of the bath spectral density and discuss the feasibility of those properties in an experiment.
\end{abstract}

\maketitle

\section{Introduction}





Ultracold atomic systems offer a platform for the engineering of matter phases with desired properties \cite{Jaksch1998,Jaksch2005,YagoMalo2024} and the quantum simulation of materials given their high resolution \cite{Gross2021, Bakr2009, Haller2015} and experimental controllability from their interaction strength and range, lattice depth, dimensionality or their temperature and, importantly, also including their dissipative coupling to their environments~\cite{Breuer2007,deVega17,Weimer2021}. In this context, atoms inside a cavity QED setup~\cite{Mivehvar2021}, offer additional versatile tools to engineer quantum many-body Hamiltonians. Here, long-range interactions as well as phonon simulation, can be engineered via coupling the atoms with a single or several cavity modes, as made possible in ring~\cite{Schmidt2014, Gangl2000, Mivehvar2018, Klinner2006} or confocal~\cite{Kollár2015, Vaidya2018} cavities, for example. Altogether, this allows to explore the emergence of a wide range of phenomena, such as frustration and glassiness~\cite{Gopalakrishnan11}, quantum liquids~\cite{Gopalakrishnan2009}, dynamical gauge fields~\cite{Ballantine17}, associative memories~\cite{Torggler17, Rotondo2018, Fiorelli2020, Marsh2021}, scrambling of information~\cite{Bentsen2019}, (dissipative) phase transitions~\cite{Baumann2010, Klinder15, Benary2022, Ferri2021}, including the Mott insulator-superfluid transition~\cite{Greiner2002} and the BCS-BEC crossover~\cite{Chen2005}, and many others.

Describing properly the dynamics of the atoms in these setups generally requires taking into account the dissipative mechanisms and the decoherence that naturally arises, altering the behavior with respect to the one of isolated systems. Furthermore, recent studies highlight the possibility of engineering noise and dissipation in an open quantum system framework to access quantum phases of matter in a way that, at times, could be more convenient than a more conventional approach of Hamiltonian Engineering~\cite{Jin2016, Lee2013}. In particular, it has been shown ~\cite{tindall} that long-range order robust against heating could be engineered in a Hubbard model by adopting a symmetry-based dissipative mechanism of the Lindblad form~\cite{Lindblad1976, Gorini1976}. The final state of the above mentioned dynamics displays so-called $\eta$-pairing, predicted first by Yang~\cite{Yang1989}. The $\eta$-state is characterized by off-diagonal long-range order (ODLRO) in the correlations between a local pair of fermions (spin up, spin down on the same site) and a holon. Importantly, this mechanism is behind finite-momentum fermionic superfluid states that are also of interest in condensed matter physics \cite{Agterberg_2020,Zhao_2023,Zhang_2024} due to their exotic properties and the emergence of associated vestigial phases.

In this paper, we study how the dissipative engineering mechanism of ODLRO via $\eta$-pairing developed in~\cite{tindall} generalizes to the case where the fermions on the lattice possess spin exchange coupling, i.e., in the tUJ model, which exhibits a different symmetry structure and which could be implemented in multimode cavity QED~\cite{Colella2018, Schlawin2019, Camacho2017, Davis2019, Norcia2018, Baier2018, Mazurenko2017}. In addition, we investigate the effect of a more general form of dissipation via the use of a Redfield master equation~\cite{Redfield1957} that, while still requiring a weak system-bath coupling, goes beyond the Lindblad dissipation considered in~\cite{tindall}. We thus show that the dissipation engineering of superfluidity still holds for the tUJ model and depends crucially on properties of the spectral density of the bath coupled to the tUJ model when one considers more general system-bath coupling than the idealized Lindblad limit.

Our paper is organized as follows. In Section II, we present the tUJ model and its symmetries. In Section III, we summarize the dissipation engineering mechanism of~\cite{tindall} allowing for stabilizing robust superfluidity against heating and apply it to the tUJ model. In Section IV, we revisit the dissipative mechanism beyond Lindblad via the use of a Redfield master equation, and study the influence of the frequency cutoff of the spectral density and its symmetry around the frequency axis. In Section V, we elaborate on a mechanism to explain our results. In Section VI, we discuss possible experimental implementations of our results. Finally, in Section VII, we conclude and provide a few possible perspectives of our work.

\section{Model and its symmetries} 

We consider the tUJ model given by the following Hamiltonian
\begin{equation}
\begin{split}
H=&-t \sum_{i,\sigma}\left(c_{i \sigma}^{\dagger} c_{i+1 \sigma}+\text { h.c. }\right)\\
&-U \sum_{i} n_{i \uparrow} n_{i \downarrow}+J \sum_{i}\left(s_{i}^{+} s_{i+1}^{-}+\text {h.c. }\right),
\end{split}
\label{eq:tuj}
\end{equation}
where $c_{i \sigma}, c_{i \sigma}^{\dagger}$ are fermionic operators in second quantization, $i$ and $\sigma$ label lattice sites and spin degrees of freedom (DOF). These define the number $n_{i,\sigma}=c_{i \sigma}^{\dagger}c_{i \sigma}$, and spin ladder $s_{i}^{+/-}=c_{i \uparrow / \downarrow}^{\dagger}c_{i \downarrow/\uparrow}$ operators, with t the tunneling amplitude, $U$ the onsite interaction and $J$ the spin exchange coupling, that we chose, without loss of generality, to be planar, requiring 
an axial magnetic anisotropy.
The state preparation scheme presented in this work is symmetry-based. Thus, we now summarize the model properties emerging from its symmetries and highlight the connections with other similar Hamiltonians where dissipative preparation schemes have been performed. To do so, we first consider useful properties of the Hubbard Hamiltonian \cite{tindall,Tindallthesis} which corresponds to our model in the limit $J\rightarrow 0$.

The Hubbard Hamiltonian 
\begin{equation}
H_{HB}= t \sum_{i,\sigma}\left(c_{i \sigma}^{\dagger} c_{i+1 \sigma}+\text { h.c. }\right)-U \sum_{i} n_{i \uparrow} n_{i \downarrow},
\label{eq:hubbard}
\end{equation}
i.e., Eq.~(\ref{eq:tuj}) with $J=0$, has, at half filling, the useful symmetry structure of a double $SU(2)$ symmetry group. The first $SU(2)$ symmetry of this Hamiltonian can be expressed via the global representation of the spin number raising and lowering operators $S^{\pm}$:
\begin{equation}
\begin{aligned}
&S^{z}=\sum_{i} s_{i}^{z}=\sum_{i}\left(n_{\uparrow, i}-n_{\downarrow, i}\right), \\
&S^{+}=\sum_{i} s_{i}^{+}=\sum_{i} c_{i, \uparrow}^{\dagger} c_{i, \downarrow}, \\ 
&S^{-}=\sum_{i} s_{i}^{-}=\sum_{i} c_{i, \downarrow}^{\dagger} c_{i, \uparrow},
\end{aligned}
\end{equation}
where $S^{z}$ is the total magnetization. These operators form a $SU(2)$ algebra and commute with the Hamiltonian, i.e.,
\begin{align}
&[H_{HB}, S_z] = 0, \\
&[H_{HB}, S_\pm] = 0, \\
&[H_{HB}, S_+S_-] = 0,
\end{align}
which thus possesses a permanent spin $SU(2)$ symmetry when defined over an arbitrary lattice. This symmetry is a manifestation of the spin-rotational invariance that characterizes the model. 

The second $SU(2)$ group corresponds to the particle-hole symmetry, via the global representation of the $\eta$ number and ladder operators:
\begin{equation}
\begin{aligned}
&\eta^{z}=\sum_{i} \eta_{i}^{z}=\sum_{i}\left(n_{\uparrow, i}+n_{\downarrow, i}-1\right), \\
&\eta^{+}=\sum_{i} \eta_{i}^{+}=\sum_{i} f(i) c_{i, \uparrow}^{\dagger} c_{i, \downarrow}^{\dagger},\\
&\eta^{-}=\sum_{i} \eta_{i}^{-}=\sum_{i} f(i) c_{i, \downarrow} c_{i, \uparrow},
\end{aligned}
\end{equation}
where $f(i)$ is a real-valued function depending on the specific lattice structure, which is $f(i)=(-1)^{i}$ for 1D bi-partite lattice. The $\eta_\pm$ operators act on empty sites (so-called holons) and on doubly-occupied sites (doublons). The $\eta$ operators form a $SU(2)$ algebra and commute with the Hamiltonian (but only at half-filling regarding the $\eta_\pm$), i.e.,
\begin{align}
&[H_{HB}, \eta_z] = 0, \\
&[H_{HB}- U N/2, \eta_\pm] = 0, \\
&[H_{HB}, \eta_+ \eta_-] = 0,
\end{align}
where $N = \sum_{i,\sigma} n_{i\sigma}$ is the number operator and $U/2$ represents the chemical potential giving rise to the particule-hole symmetry $c_{i\sigma} \to (-1)^i c^\dagger_{i\sigma}$. 

If we consider the addition of a spin exchange interaction between neighbouring fermions (XY-type anisotropy), as it is present in the tUJ model~(\ref{eq:tuj}), the $SU(2)$ spin symmetry of the system does not hold anymore, i.e.,
\begin{align}
&[H_{J}, S_z] = 0, \\
&[H_{J}, S_\pm] \neq 0, \\
&[H_{J}, S_+S_-] \neq 0.
\end{align}
As only the commutation $[H_J, S^z]$ remains zero, the system is only characterized by a $U(1)$ spin symmetry related to spin rotations about the z-axis. However,  the $SU(2)$ charge symmetry in the half-filling case remains valid, i.e.,
\begin{align}
&[H_{J}, \eta_z] = 0, \\
&[H_{J}- U N/2, \eta_\pm] = 0, \\
&[H_{J}, \eta_+ \eta_-] = 0.
\end{align}
Hence, while the Hubbard model is characterized by a $SU(2)$ spin and charge symmetry at half-filling, the tUJ model is only characterized by a \textbf{$U(1)$ spin symmetry} and \textbf{$SU(2)$ charge symmetry}.

In the following section, we describe how the underlying double symmetry structure present both in the Hubbard and tUJ models will be crucial for the realization of the fermionic superfluidity through the symmetry-based dissipative mechanism~\cite{tindall}.

\section{Dissipative state preparation scheme}\label{sec:3}

\begin{figure}[tb]
    \centering
\includegraphics[width=\columnwidth]{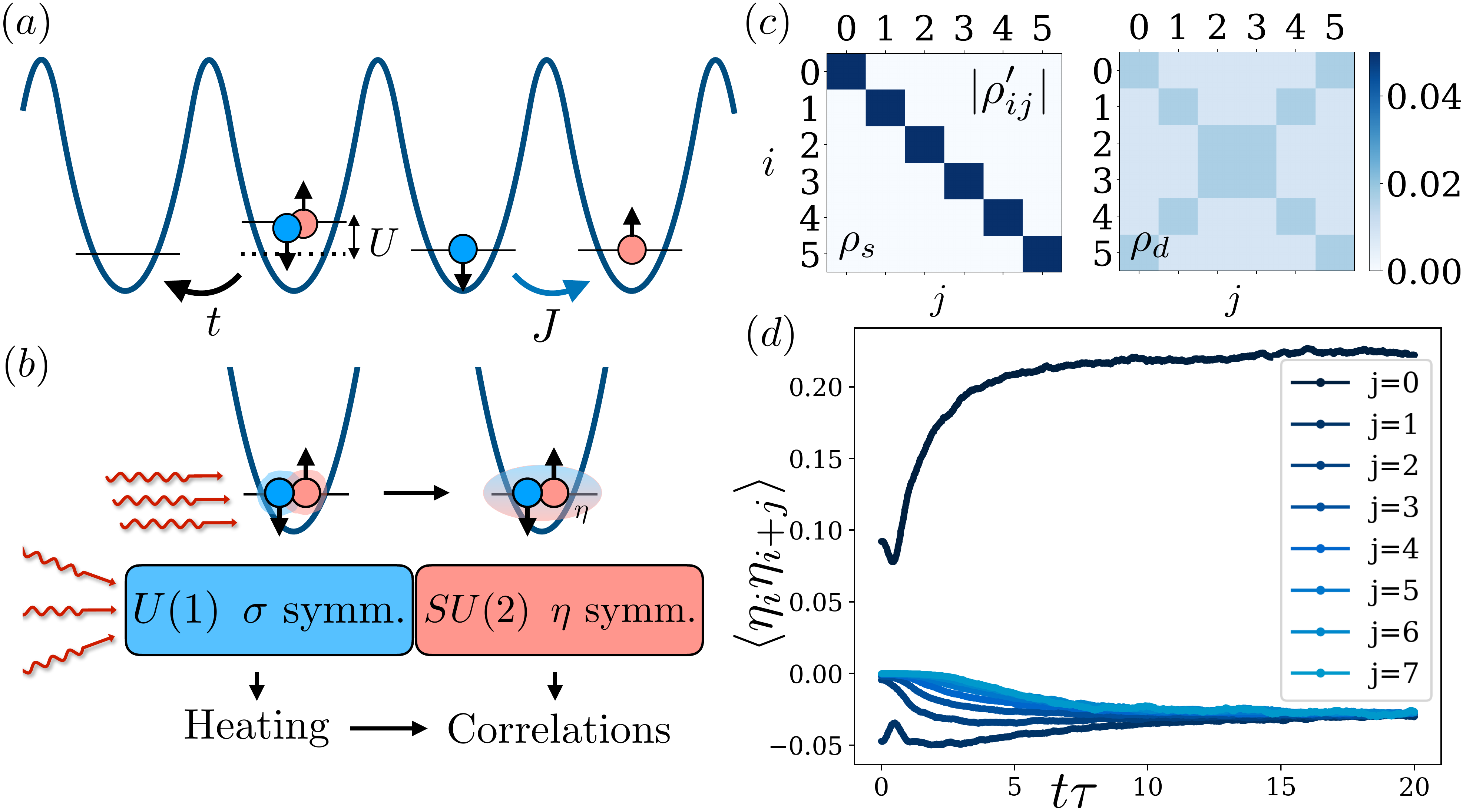}
    \caption{The concept. Superfluid formation in the tUJ model under dissipative dynamics. (a) Scheme of a fermionic chain whose dynamics are governed by the tUJ model: fermions can tunnel between neighbouring sites with rate t and experience an energy offset $U$ when two opposite spins are in the same site, and spin exchange interaction with nearest-neighbours at rate $J$. (b) Dissipative engineering of fermionic state. Pictorial representation of the dual $SU(2)+ U(1)$ symmetry in the tUJ model and the dissipative heating mechanism targeting the spin sector and leading to correlations in the $\eta$ sector. (c) Projections of the steady-state density matrix in the spin $\rho_s=P_s\rho P_s$ and charge $\rho_d= P_d \rho P_d$ sectors for $M=6$ particles at half-filling, after a quench with onsite coupling strength $U=4 t$ and dissipative rate $\gamma=2$. For the formal definition of the projector operators $P_{s,d}$, see \cite{Tindallthesis,tindall}.
    (d) Evolution of the two-point superfluid pairing correlation function $\langle\eta^+_i\eta^-_{i+j}\rangle$ for varying distances $j$ as in the legend, with $t=1, U=4t$ and $U/J=-0.1$ at half-filling. Here $M=8$ sites. Time and energy units are in units of the tunneling rate t unless stated otherwise. The pairing correlations reach the steady state approximately after $t\tau\sim 13$, with a robust condensate fraction as large as $ \approx 27\%$. The simulation is performed with $N=2000$ trajectories and timestep $\delta \tau=0.01$.}
    \label{res:8sites}
\end{figure}

Our aim is to investigate the possibility of preparing a state with long-range order in the charge sector of the tUJ model by engineering the dissipative dynamics in our system, as in~\cite{tindall}, to target a single symmetry sector so to generate a highly-correlated state in another one. To achieve this, the idea is to couple the system governed by the Hamiltonian in Eq.~(\ref{eq:tuj}) to an environment that induces spin-dependent dephasing. This could be realized by e.g. immersion of the fermionic species in a bosonic one~\cite{Klein2007}. In this section, we first model such a dissipative preparation scheme in the simplest weak system-reservoir coupling limit, as in~\cite{tindall}. In the next section, we study how going beyond this idealized scenario modifies the system dynamics.

Hence, let us first consider that the evolution of the reduced density matrix $\rho$ of our system is described by the GSKL (Gorini-Kossakowski-Sudarshan-Lindblad) master equation
\begin{equation}
\begin{aligned}
&\frac{\partial \rho}{\partial t}=\mathcal{L} \rho=-i[H, \rho]+ \frac{\gamma}{2} \sum_{j}\left([L_j \rho, L_j^\dagger] + [L_j, \rho L_j^\dagger]\right),\
\end{aligned}
\label{eq:lin_tindall}
\end{equation}
where $\gamma$ is a constant dissipative and $L_{j}$ the jump operators given by
\begin{equation}
    L_{j}=s_{j}^{z}=n_{\uparrow, j}-n_{\downarrow, j}.
\end{equation}

In Fig.~\ref{res:8sites}, we show that it is possible to stabilize a $\eta$-superfluid under such a GSKL dissipative dynamics for the tUJ model shown in Fig.~\ref{res:8sites}(a). The key aspect of the mechanism is pictorially illustrated in Fig.~\ref{res:8sites} (b); we see that the spin sector is driven to infinite temperature through the coupling to the external environment that induces heating, while the order in the charge sector is preserved, due to the symmetry:
\begin{equation}
[L_j, \eta^\pm] = [L_j, \eta^+\eta^-] = 0.
\end{equation}
This can be assessed via the reduced density matrix of the system in the respective symmetry sectors shown in Fig.~\ref{res:8sites}(c), where we can see that the steady state density matrix is diagonal in the spin sector, while it exhibits off-diagonal elements in the charge sector. The fact that the steady state density matrix is diagonal in the spin sector can be viewed as a consequence of the hermiticity of the jump operator. This can indeed be easily shown that $\rho \propto \mathds{1}$ is always a valid steady state of a Lindblad master equation of the form~(\ref{eq:lin_tindall}), as the map is unital. 
Finally, in Fig.~\ref{res:8sites}(d) we assess how the system relaxes to the steady state via the time-evolution of the two-point correlations $\langle \eta^+_i\eta^-_{i+j}\rangle$ for varying distances $|i-j|$. We observe that the stationary value of the correlations is finite and constant as a function of distance, highlighting the establishment of a long-range order with correlation values compatible with a condensate fraction as large as $n \approx 0.27$. Note that the existence of non-zero and constant steady-state values of $\langle \eta^+_i\eta^-_{i+j}\rangle$ is a sufficient condition for the existence of ODLRO, not a necessary and sufficient condition like the Penrose-Onsager criterion. The spreading of the $\eta$ correlations is visible by the decrease of the short-range correlations to favor the emergence of a distance-invariant long-range order. Hence, our results show that the dissipative mechanism to engineer long-range order described in~\cite{tindall} is still valid in our tUJ model despite the presence of the spin exchange energy term. This result generalizes the findings of \cite{tindall} to the paradigmatic tUJ model, that is amenable to simulate the presence od tunable-range interactions. However, implentations in experiment requires more realistic scenarios. 
In the following, we investigate how going beyond the weak-coupling limit approximation assumed above and in~\cite{tindall} reflects in the feasibility of the preparation schemes.
To this aim, we consider reservoirs with spectral structures.

\section{Dissipative preparation scheme beyond GSKL}\label{sec:4}

In this section we investigate whether the same SF state could be realized in more general scenarios, namely when the system is coupled to baths with a structured bath then the idealized case of an unstructured environment. In general, most materials and experiments are coupled to spectrally structured environments~\cite{deVega17}, and it is thus fundamental to investigate their effects on the system dynamics and properties, even perturbatively, leading to different behaviors and phenomena, notably in the context of dissipative phase transitions~\cite{Damanet19, Palacino21, Debecker23PRL, Debecker23PRA, Debecker2025}. It is therefore natural to ask the extent to which the results in Sect.\ref{sec:3} hold in the presence of more realistic colored baths.

In order to do so, we consider a more general form of the system evolution than that provided by Eq.~(\ref{eq:lin_tindall}), namely a Redfield master equation of the form
\begin{equation}\label{redfield}
    \dot\rho = - i [H, \rho] + \sum_{j}\left([\overline{L}_j(\tau) \rho, L_j^\dagger] + [L_j, \rho \overline{L}_j^\dagger(\tau)]\right),
\end{equation}
with
\begin{equation}\label{Lbarre}
    \overline{L}_j(\tau) = \int_0^\tau \alpha(s) e^{-i H s}L_j e^{i H s} \mathrm{d}s,
\end{equation}
and the zero-temperature bath correlation function
\begin{equation}\label{bcf1}
\alpha(s) = \int_0^{\infty} J(\omega) e^{- i \omega s} d\omega
\end{equation}
written in terms of the bath spectral density
\begin{equation}
    J(\omega) = \frac{\pi}{\hbar} \sum_k g_{k}^2 \delta(\omega-\omega_{k}),
\end{equation}
where $g_k$ denotes the system coupling with the mode $k$ of frequency $\omega_k$ of the bath. The Redfield master equation is completely determined from the knowledge of $H$, $L_j$ and the bath correlation function $\alpha(s)$, or equivalently the bath spectral density $J(\omega)$. One can directly see from Eq.~(\ref{Lbarre}) that (\ref{redfield}) reduces to (\ref{eq:lin_tindall}) if the bath spectral density is completely flat, and either one consider neglecting the imaginary part or extending the lower bound of the integration to $-\infty$, which indeed yields a bath correlation function of the form $\alpha(s) \propto \gamma \delta(s)$. By contrast, if the bath has a non-flat spectral density, the bath correlation function departs from such an instantaneous decaying form. Hence, the jump operators $L_j$ appearing in the integral of Eq.~(\ref{Lbarre}) have time to rotate according to the system Hamiltonian before the decay of the correlation function, which makes the dissipative terms in the master equation dependent on the system spectrum and eigenstates. Already from that fact, we can foresee this is likely to break the hermiticity of the jump operator, preventing the infinite temperature state to appear.

Here, we investigate the effects of various kind of spectral densities $J(\omega)$ on the dissipative phase transition discussed in Sec.~III. To interpolate between the weak-coupling limit and Redfield cases, we extend the lower bound of the integration appearing in Eq.~(\ref{bcf1}) to $- \infty$ and consider
\begin{equation}
\alpha(s) = \int_{-\infty}^{\infty} J(\omega) e^{- i \omega s} d\omega,
\end{equation}
so that the limit of $J(\omega) \to \gamma/(2\pi) \,\forall \omega$ (i.e., flat spectral density) recovers Eq.~(\ref{eq:lin_tindall}). As we will see below, the main qualitative features of our results do not depend on the exact shape of the spectral density but rather on its (i) frequency cutoff and (ii) symmetry with respect to the zero-frequency axis. Thus, for the sake of conciseness, we present the case of an effective linear spectral density, while other cases are summarised in Appendix A. We solve the master equation~(\ref{redfield}) in its Bloch-Redfield 
 form, employing the QuTip python library~\cite{qutip}. Numerical details can be found in Appendix B.

\subsection{Impact of the frequency cutoff of the spectral density}

\begin{figure}[tb]
    \centering
\includegraphics[width=\columnwidth]{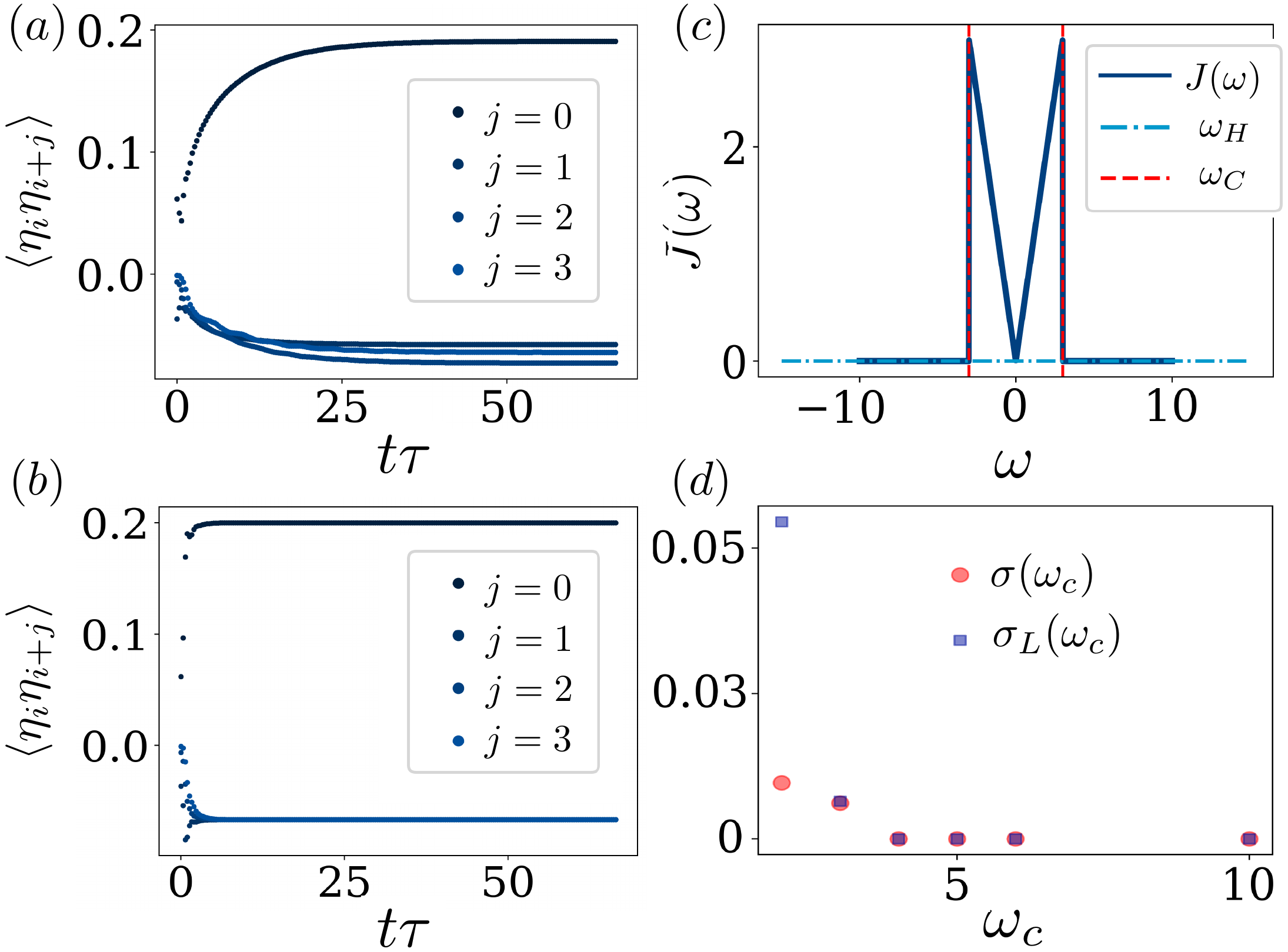}
    \caption{ Analysis of dissipative dynamics in the tUJ model with spctral structured baths. (a) Time evolution of $\langle \eta^+_{i} \eta^{-}_{i+j}\rangle$ under the dissipative dynamics in  Eq. \ref{redfield} beyond GSKL with an effective linear spectral density specified in (c) with cutoff frequency $\omega_c = 3/t$. (b) Time evolution of $\langle \eta^+_{i} \eta^{-}_{i+j}\rangle$ for a system with the same linear spectral density but with a cutoff $\omega_c = 10/t$. (c) Corresponding form of the bath correlation function $J(\omega)=|\omega|$ for an exemplary cutoff $\omega_c=3/t$ in the interval of the Hamiltonian eigenfrequencies $\omega_H$. Simulation with $M=4$ sites for $U=4t, J=-0.1U$. (d) Comparison of $\sigma(\omega_c)$ and $\sigma_L$ as a function of the cutoff $\omega_c$, where $\sigma(\omega_c) = 0$ implies a convergence to ODLRO, while $\sigma_L = 0$ implies a convergence of the off-diagonal elements to the value reached in the fully Markovian limit.}
    \label{fig:Ohm}
\end{figure}

 First, let us highlight the relevance of the frequency cutoff of the spectral density for the state preparation protocol to be effective. To understand this mechanism, we chose  $J(\omega) \propto |\omega| ,\,\textrm{for}\, -\omega_c \leq \omega \leq \omega_c $ where $\omega_c$ is a low frequency cutoff as compared with the differences between eigenfrequencies of the Hamiltonian, i.e., the transition energies of the system. 
In Fig.~\ref{fig:Ohm}(a), we observe that with a sufficiently low frequency cutoff we find no value of the bath parameters for which we obtain a steady state with constant $\eta$ correlations $\langle \eta_i \eta_{i+j} \rangle$ over distance. By contrast, for a larger cutoff frequency $\omega_c\sim10$, which encompasses all possible frequency differences between the system Hamiltonian eigenfrequencies, we observe in Fig.~\ref{fig:Ohm}(b) long-range order in the long-time limit for the two-point $\eta$ correlations. In the lower cutoff case, there is a lack of off-diagonal long-range order in the steady state -- the state is still thermal in the spin sector -- and the $\eta$ correlations are no longer able to organize themselves in a translation-invariant state. The corresponding spectral density $J(\omega)$ used in both preparation schemes are included in Fig.~\ref{fig:Ohm}(c).
While here we include only illustrative examples, we find that this discrepancy from ODLRO becomes smaller with increasing cutoff. 
We observe that both the time needed to converge to the steady state and the possibility to create long-range order depend strongly on the chosen cutoff frequency to study the convergence. Thus, we introduce a quantifier for the variation in off-diagonal correlations, defined as
\begin{equation}\label{sigmaomega}
\sigma(\omega_c) = \sqrt{\frac{1}{N}\sum_{j=1}^{N} \left(\langle \eta^+_{i} \eta^{-}_{i+j}\rangle - \mu(\omega_c)\right)^2},
\end{equation}
where $\mu(\omega_c) = \sum_{j=1}^{N} \frac{1}{N} \langle \eta^+_{i} \eta^{-}_{i+j}\rangle $ represents the mean value of the correlations, depending on the cutoff frequency $\omega_c$ through $\langle \eta^+_{i} \eta^{-}_{i+j}\rangle $. Here, we choose the central site $i=N/2+1$ without loose of generality, as the results are independent of the chosen $i$. We evaluate this quantifier for the case of an effective linear spectral density as a function of $\omega_c$ in Fig.~\ref{fig:Ohm}(d). We include as well the comparison with the fully Markovian case Eq.~(\ref{eq:lin_tindall}) via 
\begin{equation}\label{sigmaL}
    \sigma_L(\omega_c)=\sqrt{\frac{1}{N}\sum_{j=1}^{N} \left(\langle \eta^+_{i} \eta^{-}_{i+j}\rangle - \mu_L\right)^2}.
\end{equation} 
Here, $\mu_L$ corresponds to the mean value of the ODLRO correlations in the Lindblad case, allowing to quantify how $\mu(\omega_c)$ departs from $\mu_L$. From these quantifiers, we confirm that while the support of the spectral density, determined by the frequency cutoff $\omega_c$, increases and encompasses more transition energies of the system Hamiltonian, the long-range order is developed, as characterized by $\sigma(\omega_c)\rightarrow 0$. We also notice that while $\sigma_L(\omega_c) \neq 0$, the  mean value of the ODLRO correlations in the Lindblad case does not correspond to the one of the Redfield case.

\begin{figure}[!tb]
    \centering
    \includegraphics[width=\columnwidth]{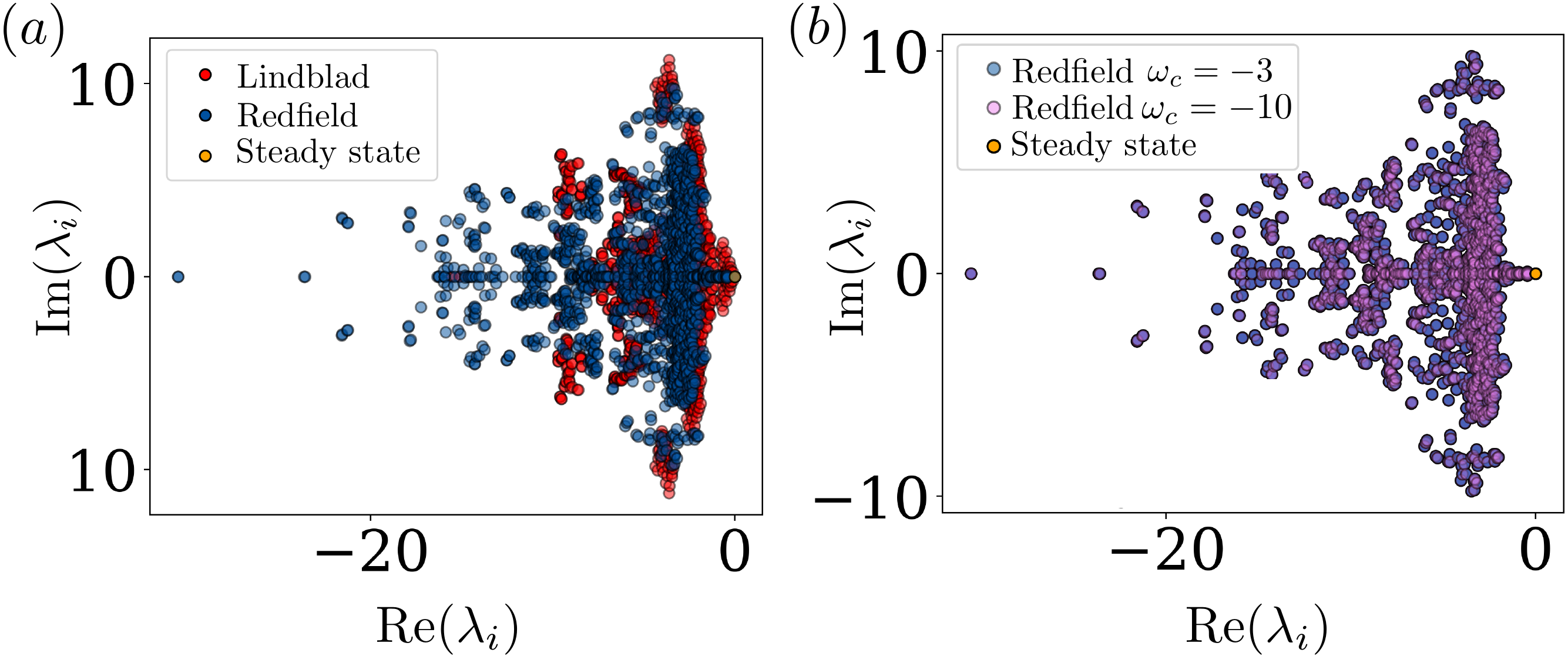}
    \caption{ (a) Eigenspace spectrum of the Redfield tensor in the case of the linear spectrum $|\omega|<3/t$ with respect to the  Liouvillian case. (b) Eigenspace spectrum of the Redfield tensor comparing an effective linear spectral density such that $|\omega|<3/t$ with respect to the Redfield tensor associated to the space spectral density but with cutoff $|\omega|<10/t$. The parameters used are the same as in Fig.\ref{fig:Ohm}.}
    \label{fig:Ohm_final}
\end{figure}

To deepen our understanding of the bath properties, we study the spectrum of the Redfield Liouvillian and compare it to one of the Lindblad Liouvillian. Our results are shown in Fig.~\ref{fig:Ohm_final}(a) for a cutoff frequency $\omega_c = 3/t$, corresponding to the evolution depicted in Fig.~\ref{fig:Ohm}(a). We observe that a small cutoff leads to the appearance of imaginary eigenvalues with small but non-zero real part 
and causes the spectrum to deviate from the standard Liouvillian case. In addition, we compare in Fig.~\ref{fig:Ohm_final}(b) the Redfield Liouvillian spectrum for the two different cutoffs investigated in Figs.~\ref{fig:Ohm}(a)-(b). For the large cutoff $\omega_c = 10/t$ (violet dots) the spectra is already comparable to the Lindblad case and we observe the ORDLO.

\subsection{Impact of the symmetry of the spectral density}

\begin{figure}[tb]
    \centering
    \includegraphics[width=\columnwidth]{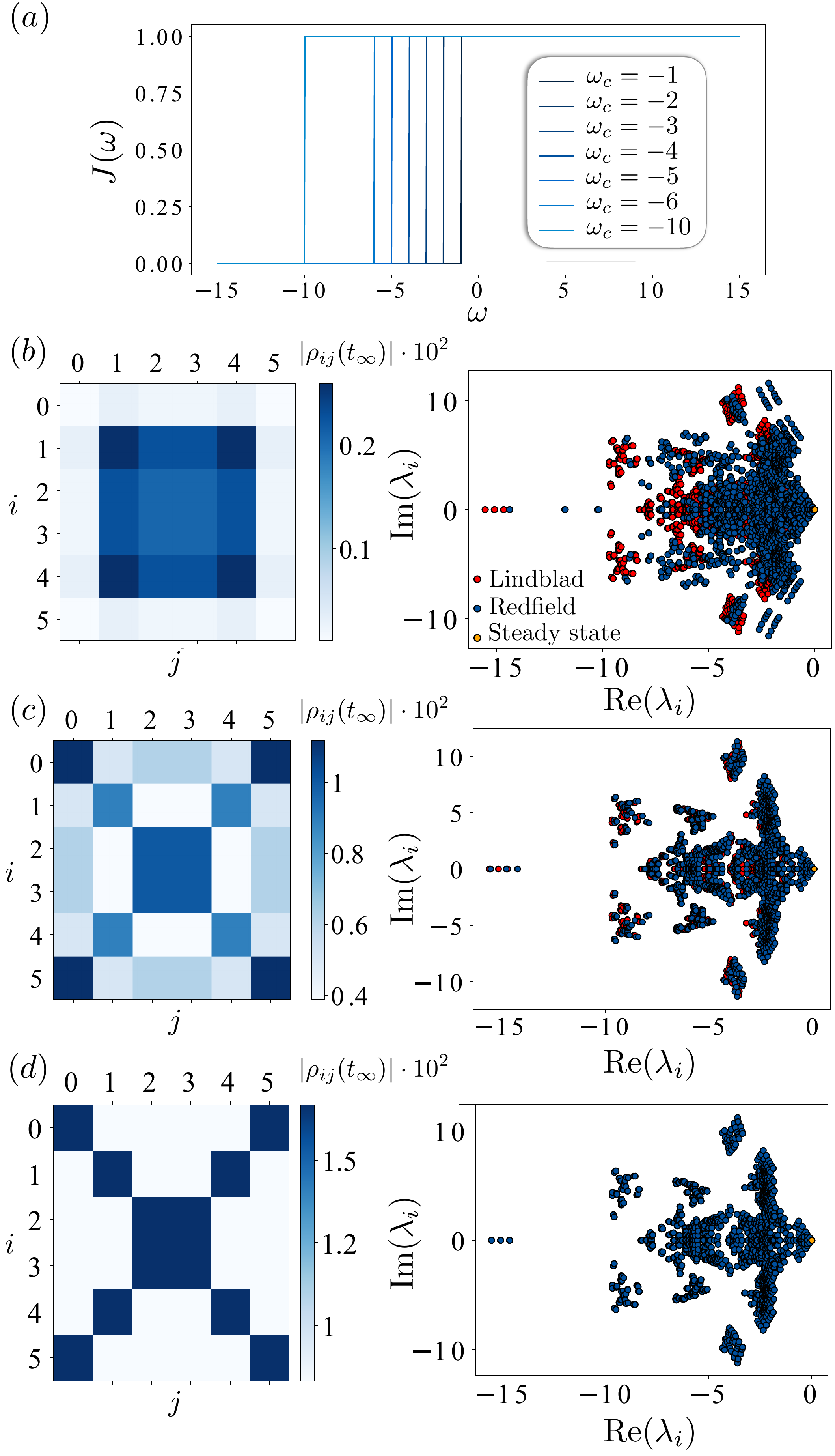}
    \caption{In Fig.~\ref{fig:dy}(a), we present the spectral density used to study the dynamics: $\theta(\omega - \omega_c)$. Figs.~\ref{fig:dy}(b)-(d) display the projections of the final state density matrix in the charge sector for cutoffs $\omega_c = \{-1, -5, -10\}$ (right column), respectively. The associated Redfield tensor eigenspectra (blue dots) are overlaid with the Liouvillian eigenspectra (red dots)  (left column). This allows us to trace the evolution of correlations between doublons, from their absence to the emergence of off-diagonal long-range order (ODLRO) in the sector.} The parameters used are the same as in Fig.\ref{fig:Ohm}.
    \label{fig:dy}
\end{figure}

We now turn to the study of the impact of the spectral density symmetry with respect to the zero-energy axis. To this aim, we use by using a step-like spectral density represented in Fig.~\ref{fig:dy}(a), where the step occurs at $\omega_c$ a parameter, that we vary from $0$ to large negative values. For this analysis, we study the variation of the reduced density matrix projected onto the target $\eta$ sector as a function of the position of the step in the spectral density. For a low (negative) cutoff [panel (b) of Fig.~\ref{fig:dy}], the reduced $\eta$ state does not display translational invariant order while for a large (negative) cutoff (panel (d) of Fig.~\ref{fig:dy}), the long-range order is established and it is possible to stabilize the $\eta$-superfluid. Intermediate frequency ranges (panel (c)) show both the initial creation of correlation at the boundaries and the redistribution of correlations on the entire chain as we vary $\omega_c$. The increasing overlap between the Lindblad and Redfield Liouvillian spectra as the cutoff frequency becomes larger in modulus (right plots in panels (b-d)) confirms the convergence of the model to the known Lindblad scenario associated with the stable $\eta$-superfluid.

We interpret the lack of identical $\eta$-correlations at low $\omega_c$ as a consequence of the asymmetry of the spectral density on the transition energy range of the Hamiltonian. Indeed, in this case, the rotated jump operator in Eq.~(\ref{Lbarre}) becomes non-Hermitian, which breaks the unitality of the quantum map, that is a condition to obtain equal $\eta$-correlations~\cite{tindall}. As the cutoff becomes large in modulus, the symmetry over the relevant range of frequencies of the spectral density is recovered, and so the hermiticity of the rotated jump operator and the unitality of the map. These outcomes can be extended to other spectral structures, as discussed in App. A.

\section{Discussion}

From the analysis in Sect.\ref{sec:4} we extract two essential indications: (i) the spectral density has to be symmetrical respect to the zero-energy axis for the convergence to ODLRO to occur and (ii) the cutoff of the spectral density is required to be large enough so that it covers, symmetrically, the relevant transition energy range determined by $H$. Overall, this can be viewed as the requirement that the bath must be able to connect the eigenstates of the system to the target correlated state via relevant dissipative channels. Closing some dissipative channels via the vanishing of the spectral density can thus be detrimental for the emergence of the target state. 

Note that the observed breaking of the emergence of identical $\eta$-correlations is not due to the breaking of the strong symmetry condition (which was studied in~\cite{tindall} via the addition of Lindblad dissipative terms that explicitly break it). The fact that going beyond Lindblad does not break strong symmetries has actually already been predicted in~\cite{Buca2019} under general assumptions. In the case of our Redfield master equation, one can indeed compute $
[\overline{L}_j(\tau), \eta^\pm]$ and $[\overline{L}_j(\tau), \eta^+\eta^-] = 0$ and show that they remain zero for both asymmetric spectral density or low cutoff frequency. However, interestingly, our results show that even without breaking the strong symmetry, going beyond Lindblad can still alter the steady state. While this does not prevent the emergence of $\eta$-correlations, it hinders their convergence towards an identical value.
This information suggests that the ODLRO we aim to prepare under more general dissipation conditions really needs to be engineered around a Lindblad-type description. Specifically, states asymptotically converging to the target state can be achieved, requiring longer preparation times constrained by the conditions on the bath structure, these findings open up to the impassibility of realizing a SF state in some realistic experiments.

\section{Experimental Implementation}

1D fermionic models can be realized by tightly trapping ultracold fermionic gas of atoms in optical lattices along the longitudinal axis of a cavity QED~\cite{Colella2018, Schlawin2019, Camacho2017, Lucchesi2019, Baier2018, Mazurenko2017}, where the spin-exchange interactions of the $t UJ$ models could be implemented elaborating e.g. on~\cite{Davis2019, Norcia2018}. Multi-mode cavity setups could also be useful for controlling the range of interactions between the particles~\cite{Vaidya2018}.

Regarding the dissipation, previous proposals~\cite{tindall} for spin-dependent dephasing have been inspired in reproducing phenomenologically spin fluctuation theory~\cite{Kruchinin2010} via superfluid immersion~\cite{Daley2004,Bruderer2007} tuning appropriately the corresponding scattering lengths via Feshbach resonances~\cite{Chin2010}. Alternative approaches could make use of spin-dependent potentials, as e.g., in \cite{Mandel2003,Yang2019}, in order to induce spin-dependent coupling to the environment in a fully-optical setup. We mention the proposal in \cite{Bernier2013}, where it is shown that it is possible to engineer pair coherence over long-distances through incoherent local environments using a probing beam tuned exactly between the spin states, in a protocol inspired by phase-contrast imaging \cite{Shin2006}. 

The development of these schemes in cold-atom platforms, would provide a new testing ground for current relevant models in condensed matter.
Indeed, the superfluid state here prepared dissipatively corresponds to a superconducting state that breaks time reversal symmetry and it is characterized by finite momentum Cooper pairing \cite{Agterberg_2020}, in a condensed matter context these states are exotic with the only experimental observation, in absence of external magnetic field, reported in $EuRbFe_4As_4$ \cite{Zhao_2023}. Interestingly, in real materials this state has been theorized in the presence of magnetic fluctuations, in particular driven by the presence of altermagnetism \cite{Zhang_2024}.
Thus, the possibility of realizing such a state in an ultracold atom platform would give access to the study of its properties and, potentially, to the understanding of its governing mechanism by engineering different kinds of "spin baths" in a controlled experimental setup.
Therefore employing the ultracold atom platform as a quantum simulator for a condensed matter problem.

\section{CONCLUSION}

In this work, we have shown that it is possible to prepare a state with finite momentum by means of dissipation in a tUJ model compatible with current fermionic multimode cavity QED setups. We have analyzed the importance of the Markovian limit in the state preparation, which requires a bath that has complete overlap of the system energy spectrum and is symmetric around low frequencies. We have then identified the conditions for the preparation scheme in the presence of dissipation preserving the target symmetry, but not following the full Lindblad description. Thus, we can conclude that this state preparation requires dissipation engineering around a Lindblad dynamics in the relevant frequency range. Open questions remain regarding the effects of other forms of non-Markovianity and the role of the bath properties in more general dissipative state preparation proposals based on symmetries, that we leave for future work. 

\begin{acknowledgments}

We wish to acknowledge the support of Benjamin Lev, Jonathan Keeling and Peter Kirton for fruitful discussions during this project. M.L.C. acknowledges support from the National Centre on HPC, Big Data and Quantum Computing - SPOKE 10 (Quantum Computing) and received funding from the European Union NextGenerationEU - National Recovery and Resilience Plan (NRRP) – MISSION 4 COMPONENT 2, INVESTMENT N. 1.4 – CUP N.
I53C22000690001. J.Y.M. and M.L.C. were supported by the European Social Fund REACT EU through the
Italian national program PON 2014-2020, DM MUR 1062/2021. M.L.C. also acknowledges support from the
project PRA2022202398 “IMAGINATION”. 
\end{acknowledgments}

\appendix

\section{Other baths}
We present here the results obtained considering different kind of spectral functions for the bath.

We focus in particular on a Lorentzian spectral density $J(\omega)=\eta {\gamma}/({\gamma^2 + (\omega -\omega_c)^2})$, that can be considered to approximate well an environment made up of a single harmonic oscillator with
frequency $\omega_c$, dissipating into the vacuum at rate $\gamma$. 

\begin{figure}[h!]
    \centering
    \includegraphics[width=0.495\textwidth]{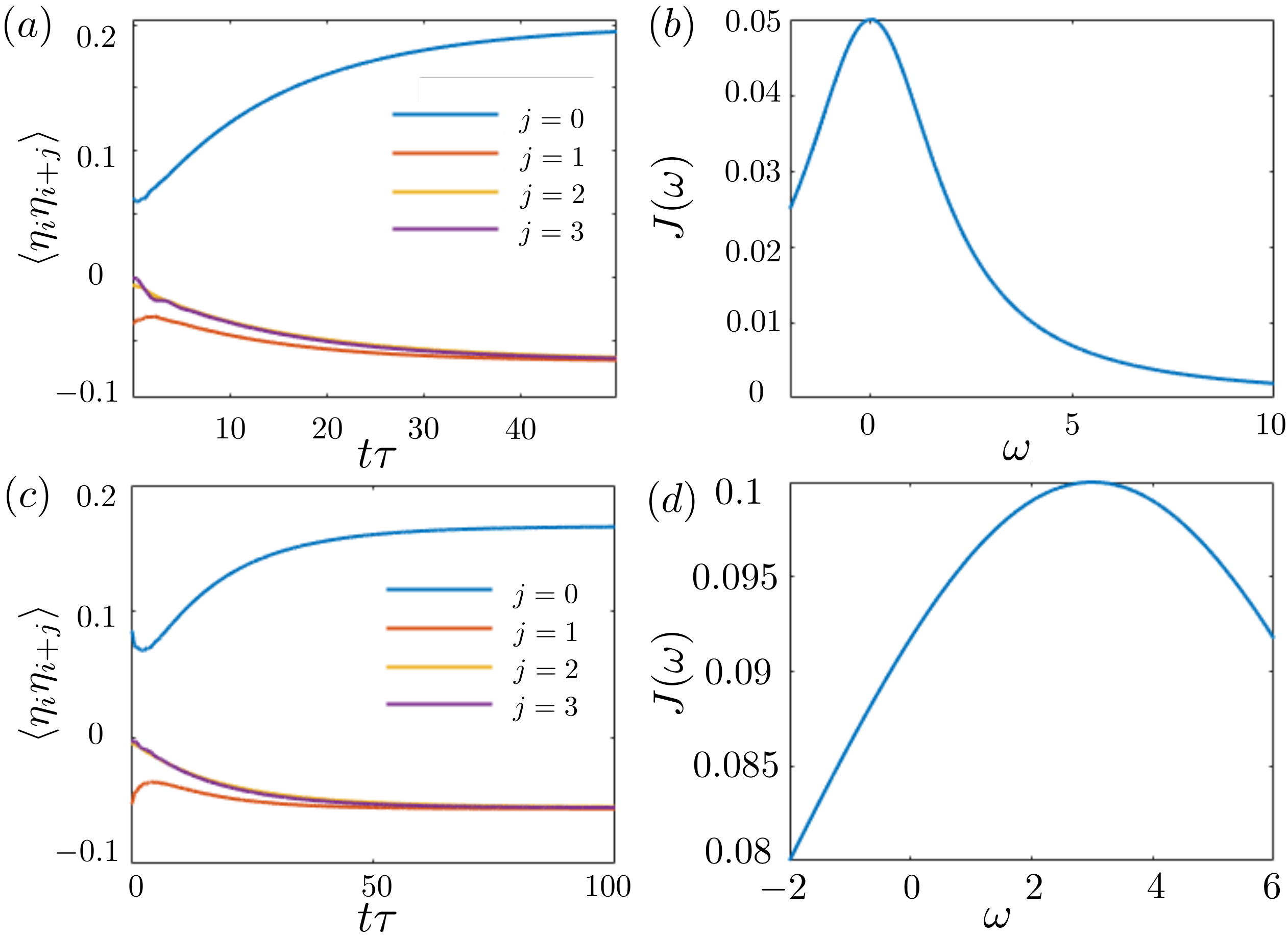}
    \caption{Time evolution of $\langle \eta^+_{i} \eta^{-}_{i+j}\rangle$ under the dissipative dynamics in  Eq. \eqref{redfield} beyond GSKL with Lorentzian spectral density $J(\omega)=\eta {\gamma}/({\gamma^2 + (\omega -\omega_c)^2})$ (see text). (a)-(c):  $\langle \eta^+_{i} \eta^{-}_{i+j}\rangle$ at different distances $j$ as in the legend. (b)-(d): the corresponding form of the bath correlation function of the form $J(\omega)$ in the interval of the Hamiltonian eigenfrequencies.   (a)-(b): $\eta=1, \gamma=2, \omega_c=0$. (c)-(d): shifted Lorentzian $\omega_c=3$,  the values of $\gamma=10$ and $\eta=1$ being tuned to ensure the same homogeneity of the spectral function in the low frequency spectrum.  We observe that in the case of a centered Lorentzian we could not find any value breaking the convergence to the steady state. Simulation with $M=4$ sites for $U=4t, J=-0.1U$.}
    \label{fig:lorentz}
\end{figure}

As we can be seen in Fig.~\ref{fig:lorentz}, moving away from $\omega_c=0$ (or if $\gamma$ is not large enough) slows down the convergence towards the steady state. 

\begin{figure}[h!]
    \centering
\includegraphics[width=0.495\textwidth]{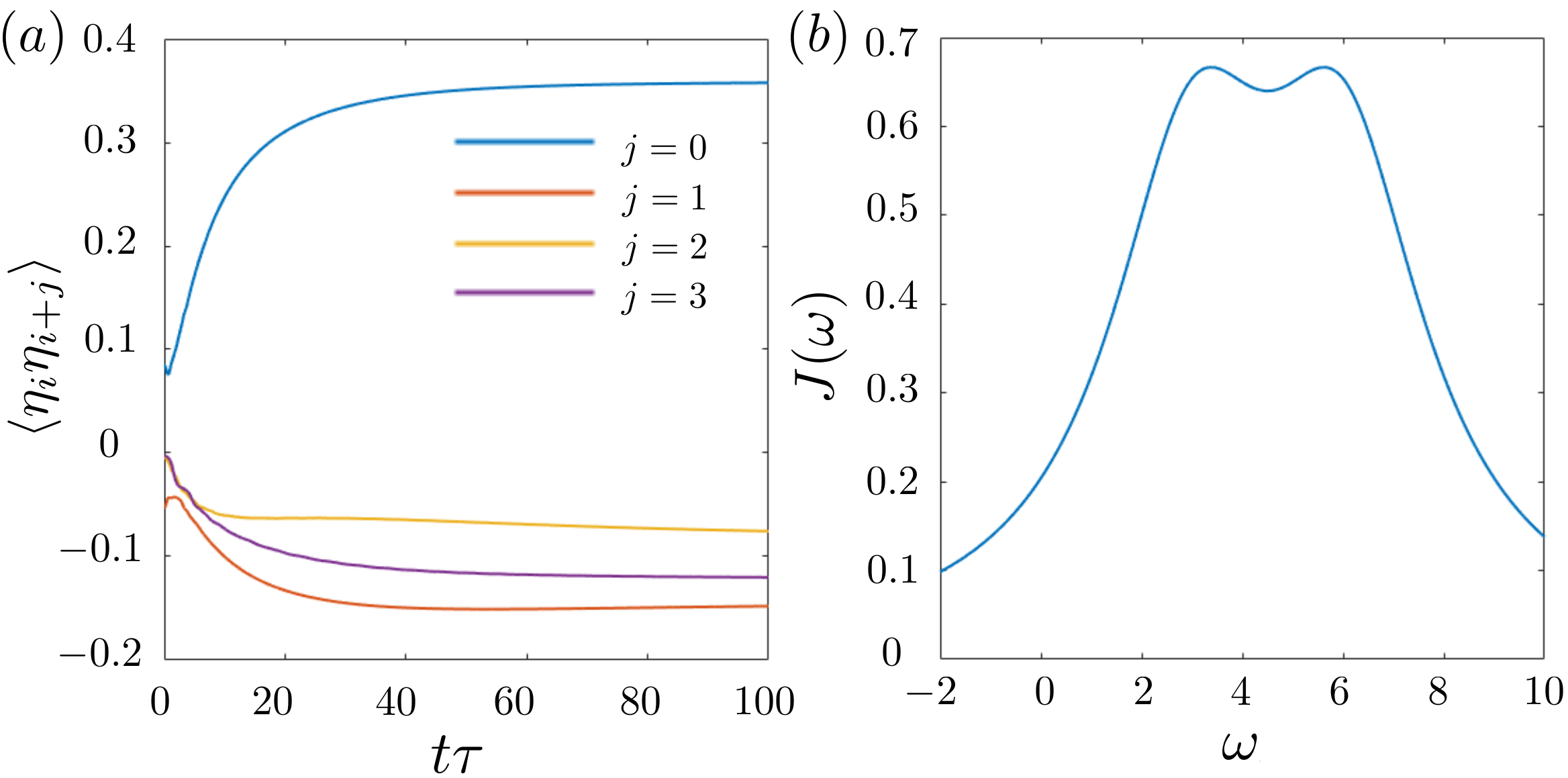}
    \caption{Time evolution of $\langle \eta^{+}_{i} \eta^{-}_{i+j}\rangle$ under the dissipative dynamics in  Eq. \eqref{redfield} with the spectral density $J(\omega)=\eta \gamma/(\gamma^2 + (\omega -\omega_c)^2) + \eta^{\prime} \gamma^{\prime}/({\gamma^\prime}^2 + (\omega -\omega^{\prime}_c)^2)$ given by the sum of two Lorentzians (see text). (a) $\langle \eta^+_{i}\eta^{-}_{i+j}\rangle$ at different distances j as in the legend; (b) the corresponding form of the bath correlation function of the form $J(\omega)$ in the interval of the Hamiltonian eigenfrequencies. The parameters values are: $\eta=1, \gamma=2, \omega_c=3, \eta^{\prime}=1, \gamma^{\prime}=2, \omega_{c}^{\prime}=6$. While it is of course possible to achieve convergence by fine tuning of the values until a sufficiently flat function is obtained, we observe that it is hard to obtain convergence also for a slightly structured bath if the frequencies involved fall in the relevant range of the Hamiltonian spectrum. Simulation with $M=4$ sites for $U=4t, J=-0.1U$.}
    \label{fig:structured}
\end{figure}

As a structured spectral density with several peaks is a generic feature of real materials, we also show results of a spectral density built as the sum of two shifted Lorentzians in Fig.~\ref{fig:structured}. As we can be seen, depending on the behavior near low frequencies, we can again drastically slow down the convergence towards the steady state. 

\begin{figure}[h!]
    \centering
    \includegraphics[width=0.495\textwidth]{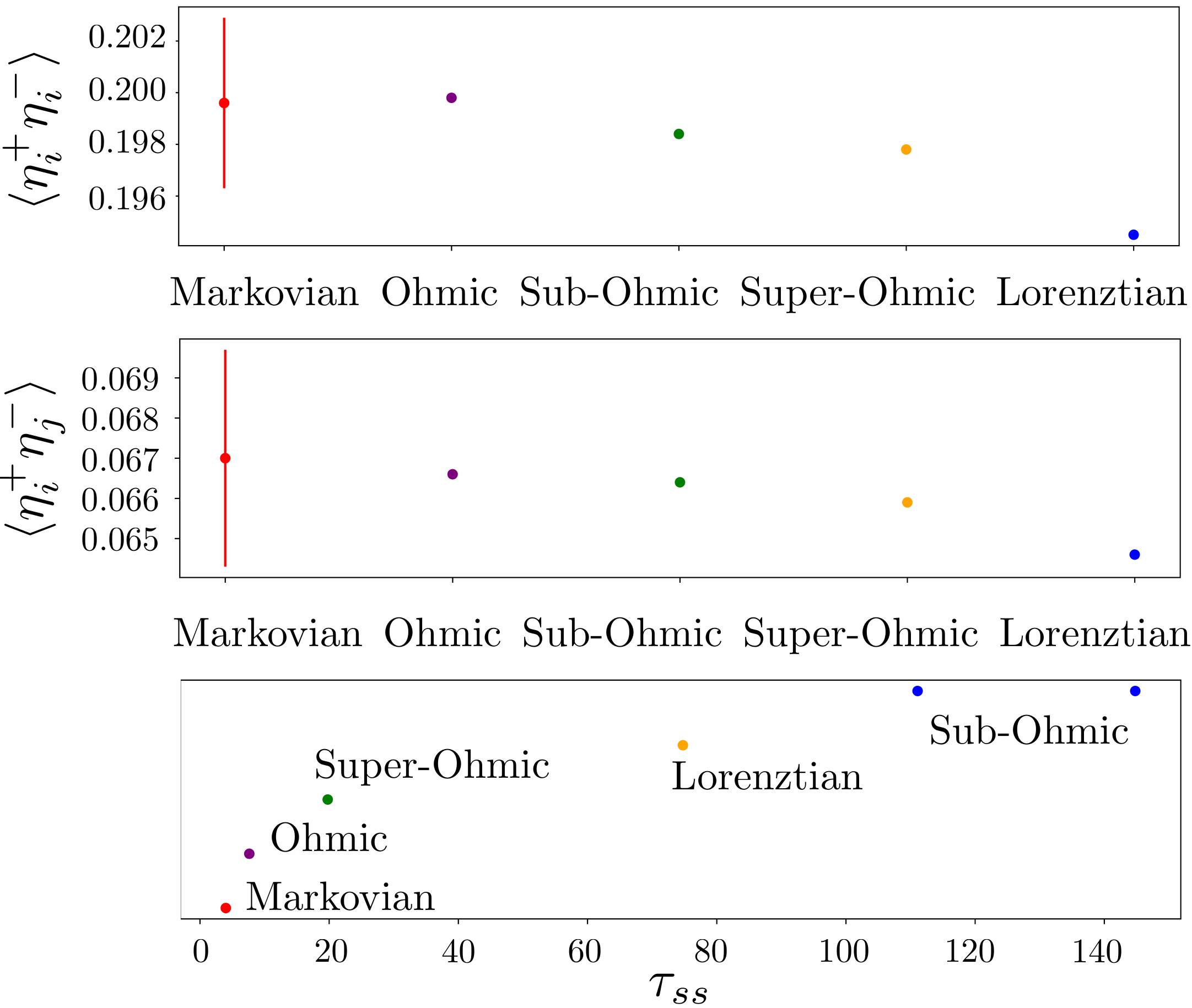}
    \caption{Summary of the beyond-GSKL analysis using different spectral densities $J(\omega)$. The values associated with each spectral density are obtained with specific and minimal parameters to achieve a steady state. For the Markovian case, calculations are performed for $M=4$ sites, consistent with the system size used for studying the Non-Markovian regime, and based on 1000 trajectories, resulting in an associated error of approximately 4\%. Top: Scatter plot of the doublon occupancy in the steady state for different spectral functions. A general decrease is observed, with the Ohmic-like (i.e., linear spectral density) case performing the best. Except for the Lorentzian case, the values obtained are consistent within error bars with the Markovian limit. Center: Off-diagonal terms of the $\eta$-correlations, showing that the Ohmic bath better reproduces the Markovian limit. For the other spectral functions, a decrease in the condensate fraction is observed, though it remains consistent with the results in the Markovian limit. Bottom: Trend of the steady-state convergence time $\tau_{ss}$ in different cases, defined as the first instant at which the final value is reached. A progressive and significant increase in the convergence time is observed as we depart from the Markovian case.}
    \label{fig:summary}
\end{figure}

In Fig.~\ref{fig:summary} we summarize the results for all the other baths~\footnote{Note: In Fig.~\ref{fig:summary} super-Ohmic, sub-Ohmic and Lorentzian spectral functions are interchangeable, depending on the functional form adopted, meaning that the values are specific and do not represent a general trend. For example, in the Lorentzian spectral function, whose width can be tuned until it is completely flat in the relevant region, we can obtain results consistent to the Markovian case. We decide here to represent the values obtained by adopting the minimal parameter in the functional form of the spectral density, so to achieve convergence without excessively deforming $J(\omega)$.} by representing on a scatter plot the mean value of the off diagonal correlations reached in $\eta$ and the time $\tau_{ss}$ at which convergence is reached by the system coupled to the corresponding bath. It is apparent that there exist regimes for each kind of bath in which we still manage to reach the steady state. These regimes are characterized by homogeneity of the bath correlation functions which are approximately symmetric in frequency at least in the neighborhood of the low frequency part of the Hamiltonian spectrum.\\

In general, the farther the functional form of the spectral density moves away from the Markovian bath, the smaller is the fraction of superfluid obtained in the final state, while the convergence time increases. The value at the steady state and the time required to get there, are both seen to strongly depend on the parameters of the spectral function. 

\newpage

\section{Numerical solution of the Bloch-Redfield equation}

In order to solve the Redfield Equation~\ref{redfield}, we use the Bloch-Redfield master equation integrator provided by the Python package QuTiP~\cite{qutip}. Given the complexity of the numerical calculation related to the exponential growth of the Bloch-Redfield tensor, which slows down the numerical integrator, we present results of a maximum of $M=8$ sites.

\bibliographystyle{apsrev4-2}
\bibliography{references}

\begin{thebibliography}{65}%
\makeatletter
\providecommand \@ifxundefined [1]{%
 \@ifx{#1\undefined}
}%
\providecommand \@ifnum [1]{%
 \ifnum #1\expandafter \@firstoftwo
 \else \expandafter \@secondoftwo
 \fi
}%
\providecommand \@ifx [1]{%
 \ifx #1\expandafter \@firstoftwo
 \else \expandafter \@secondoftwo
 \fi
}%
\providecommand \natexlab [1]{#1}%
\providecommand \enquote  [1]{``#1''}%
\providecommand \bibnamefont  [1]{#1}%
\providecommand \bibfnamefont [1]{#1}%
\providecommand \citenamefont [1]{#1}%
\providecommand \href@noop [0]{\@secondoftwo}%
\providecommand \href [0]{\begingroup \@sanitize@url \@href}%
\providecommand \@href[1]{\@@startlink{#1}\@@href}%
\providecommand \@@href[1]{\endgroup#1\@@endlink}%
\providecommand \@sanitize@url [0]{\catcode `\\12\catcode `\$12\catcode `\&12\catcode `\#12\catcode `\^12\catcode `\_12\catcode `\%12\relax}%
\providecommand \@@startlink[1]{}%
\providecommand \@@endlink[0]{}%
\providecommand \url  [0]{\begingroup\@sanitize@url \@url }%
\providecommand \@url [1]{\endgroup\@href {#1}{\urlprefix }}%
\providecommand \urlprefix  [0]{URL }%
\providecommand \Eprint [0]{\href }%
\providecommand \doibase [0]{https://doi.org/}%
\providecommand \selectlanguage [0]{\@gobble}%
\providecommand \bibinfo  [0]{\@secondoftwo}%
\providecommand \bibfield  [0]{\@secondoftwo}%
\providecommand \translation [1]{[#1]}%
\providecommand \BibitemOpen [0]{}%
\providecommand \bibitemStop [0]{}%
\providecommand \bibitemNoStop [0]{.\EOS\space}%
\providecommand \EOS [0]{\spacefactor3000\relax}%
\providecommand \BibitemShut  [1]{\csname bibitem#1\endcsname}%
\let\auto@bib@innerbib\@empty
\bibitem [{\citenamefont {Jaksch}\ \emph {et~al.}(1998)\citenamefont {Jaksch}, \citenamefont {Bruder}, \citenamefont {Cirac}, \citenamefont {Gardiner},\ and\ \citenamefont {Zoller}}]{Jaksch1998}%
  \BibitemOpen
  \bibfield  {author} {\bibinfo {author} {\bibfnamefont {D.}~\bibnamefont {Jaksch}}, \bibinfo {author} {\bibfnamefont {C.}~\bibnamefont {Bruder}}, \bibinfo {author} {\bibfnamefont {J.~I.}\ \bibnamefont {Cirac}}, \bibinfo {author} {\bibfnamefont {C.~W.}\ \bibnamefont {Gardiner}},\ and\ \bibinfo {author} {\bibfnamefont {P.}~\bibnamefont {Zoller}},\ }\href {https://doi.org/10.1103/PhysRevLett.81.3108} {\bibfield  {journal} {\bibinfo  {journal} {Phys. Rev. Lett.}\ }\textbf {\bibinfo {volume} {81}},\ \bibinfo {pages} {3108} (\bibinfo {year} {1998})}\BibitemShut {NoStop}%
\bibitem [{\citenamefont {Jaksch}\ and\ \citenamefont {Zoller}(2005)}]{Jaksch2005}%
  \BibitemOpen
  \bibfield  {author} {\bibinfo {author} {\bibfnamefont {D.}~\bibnamefont {Jaksch}}\ and\ \bibinfo {author} {\bibfnamefont {P.}~\bibnamefont {Zoller}},\ }\href {https://doi.org/https://doi.org/10.1016/j.aop.2004.09.010} {\bibfield  {journal} {\bibinfo  {journal} {Annals of Physics}\ }\textbf {\bibinfo {volume} {315}},\ \bibinfo {pages} {52} (\bibinfo {year} {2005})},\ \bibinfo {note} {special Issue}\BibitemShut {NoStop}%
\bibitem [{\citenamefont {Yago~Malo}\ \emph {et~al.}(2024)\citenamefont {Yago~Malo}, \citenamefont {Lepori}, \citenamefont {Gentini},\ and\ \citenamefont {Chiofalo}}]{YagoMalo2024}%
  \BibitemOpen
  \bibfield  {author} {\bibinfo {author} {\bibfnamefont {J.}~\bibnamefont {Yago~Malo}}, \bibinfo {author} {\bibfnamefont {L.}~\bibnamefont {Lepori}}, \bibinfo {author} {\bibfnamefont {L.}~\bibnamefont {Gentini}},\ and\ \bibinfo {author} {\bibfnamefont {M.~L.~M.}\ \bibnamefont {Chiofalo}},\ }\href@noop {} {\bibfield  {journal} {\bibinfo  {journal} {Technologies}\ }\textbf {\bibinfo {volume} {12}} (\bibinfo {year} {2024})}\BibitemShut {NoStop}%
\bibitem [{\citenamefont {Gross}\ and\ \citenamefont {Bakr}(2021)}]{Gross2021}%
  \BibitemOpen
  \bibfield  {author} {\bibinfo {author} {\bibfnamefont {C.}~\bibnamefont {Gross}}\ and\ \bibinfo {author} {\bibfnamefont {W.~S.}\ \bibnamefont {Bakr}},\ }\href {https://doi.org/10.1038/s41567-021-01370-5} {\bibfield  {journal} {\bibinfo  {journal} {Nature Physics}\ }\textbf {\bibinfo {volume} {17}},\ \bibinfo {pages} {1316} (\bibinfo {year} {2021})}\BibitemShut {NoStop}%
\bibitem [{\citenamefont {Bakr}\ \emph {et~al.}(2009)\citenamefont {Bakr}, \citenamefont {Gillen}, \citenamefont {Peng}, \citenamefont {F{\"o}lling},\ and\ \citenamefont {Greiner}}]{Bakr2009}%
  \BibitemOpen
  \bibfield  {author} {\bibinfo {author} {\bibfnamefont {W.~S.}\ \bibnamefont {Bakr}}, \bibinfo {author} {\bibfnamefont {J.~I.}\ \bibnamefont {Gillen}}, \bibinfo {author} {\bibfnamefont {A.}~\bibnamefont {Peng}}, \bibinfo {author} {\bibfnamefont {S.}~\bibnamefont {F{\"o}lling}},\ and\ \bibinfo {author} {\bibfnamefont {M.}~\bibnamefont {Greiner}},\ }\href {https://doi.org/10.1038/nature08482} {\bibfield  {journal} {\bibinfo  {journal} {Nature}\ }\textbf {\bibinfo {volume} {462}},\ \bibinfo {pages} {74} (\bibinfo {year} {2009})}\BibitemShut {NoStop}%
\bibitem [{\citenamefont {Haller}\ \emph {et~al.}(2015)\citenamefont {Haller}, \citenamefont {Hudson}, \citenamefont {Kelly}, \citenamefont {Cotta}, \citenamefont {Peaudecerf}, \citenamefont {Bruce},\ and\ \citenamefont {Kuhr}}]{Haller2015}%
  \BibitemOpen
  \bibfield  {author} {\bibinfo {author} {\bibfnamefont {E.}~\bibnamefont {Haller}}, \bibinfo {author} {\bibfnamefont {J.}~\bibnamefont {Hudson}}, \bibinfo {author} {\bibfnamefont {A.}~\bibnamefont {Kelly}}, \bibinfo {author} {\bibfnamefont {D.~A.}\ \bibnamefont {Cotta}}, \bibinfo {author} {\bibfnamefont {B.}~\bibnamefont {Peaudecerf}}, \bibinfo {author} {\bibfnamefont {G.~D.}\ \bibnamefont {Bruce}},\ and\ \bibinfo {author} {\bibfnamefont {S.}~\bibnamefont {Kuhr}},\ }\href {https://doi.org/10.1038/nphys3403} {\bibfield  {journal} {\bibinfo  {journal} {Nature Physics}\ }\textbf {\bibinfo {volume} {11}},\ \bibinfo {pages} {738} (\bibinfo {year} {2015})}\BibitemShut {NoStop}%
\bibitem [{\citenamefont {Breuer}\ and\ \citenamefont {Petruccione}(2007)}]{Breuer2007}%
  \BibitemOpen
  \bibfield  {author} {\bibinfo {author} {\bibfnamefont {H.-P.}\ \bibnamefont {Breuer}}\ and\ \bibinfo {author} {\bibfnamefont {F.}~\bibnamefont {Petruccione}},\ }\href@noop {} {\emph {\bibinfo {title} {The Theory of Open Quantum Systems}}}\ (\bibinfo  {publisher} {Oxford University Press},\ \bibinfo {year} {2007})\BibitemShut {NoStop}%
\bibitem [{\citenamefont {de~Vega}\ and\ \citenamefont {Alonso}(2017)}]{deVega17}%
  \BibitemOpen
  \bibfield  {author} {\bibinfo {author} {\bibfnamefont {I.}~\bibnamefont {de~Vega}}\ and\ \bibinfo {author} {\bibfnamefont {D.}~\bibnamefont {Alonso}},\ }\href {https://doi.org/10.1103/RevModPhys.89.015001} {\bibfield  {journal} {\bibinfo  {journal} {Rev. Mod. Phys.}\ }\textbf {\bibinfo {volume} {89}},\ \bibinfo {pages} {015001} (\bibinfo {year} {2017})}\BibitemShut {NoStop}%
\bibitem [{\citenamefont {Weimer}\ \emph {et~al.}(2021)\citenamefont {Weimer}, \citenamefont {Kshetrimayum},\ and\ \citenamefont {Or\'us}}]{Weimer2021}%
  \BibitemOpen
  \bibfield  {author} {\bibinfo {author} {\bibfnamefont {H.}~\bibnamefont {Weimer}}, \bibinfo {author} {\bibfnamefont {A.}~\bibnamefont {Kshetrimayum}},\ and\ \bibinfo {author} {\bibfnamefont {R.}~\bibnamefont {Or\'us}},\ }\href {https://doi.org/10.1103/RevModPhys.93.015008} {\bibfield  {journal} {\bibinfo  {journal} {Rev. Mod. Phys.}\ }\textbf {\bibinfo {volume} {93}},\ \bibinfo {pages} {015008} (\bibinfo {year} {2021})}\BibitemShut {NoStop}%
\bibitem [{\citenamefont {Farokh~Mivehvar}\ and\ \citenamefont {Ritsch}(2021)}]{Mivehvar2021}%
  \BibitemOpen
  \bibfield  {author} {\bibinfo {author} {\bibfnamefont {T.~D.}\ \bibnamefont {Farokh~Mivehvar}, \bibfnamefont {Francesco~Piazza}}\ and\ \bibinfo {author} {\bibfnamefont {H.}~\bibnamefont {Ritsch}},\ }\href {https://doi.org/10.1080/00018732.2021.1969727} {\bibfield  {journal} {\bibinfo  {journal} {Advances in Physics}\ }\textbf {\bibinfo {volume} {70}},\ \bibinfo {pages} {1} (\bibinfo {year} {2021})},\ \Eprint {https://arxiv.org/abs/https://doi.org/10.1080/00018732.2021.1969727} {https://doi.org/10.1080/00018732.2021.1969727} \BibitemShut {NoStop}%
\bibitem [{\citenamefont {Schmidt}\ \emph {et~al.}(2014)\citenamefont {Schmidt}, \citenamefont {Tomczyk}, \citenamefont {Slama},\ and\ \citenamefont {Zimmermann}}]{Schmidt2014}%
  \BibitemOpen
  \bibfield  {author} {\bibinfo {author} {\bibfnamefont {D.}~\bibnamefont {Schmidt}}, \bibinfo {author} {\bibfnamefont {H.}~\bibnamefont {Tomczyk}}, \bibinfo {author} {\bibfnamefont {S.}~\bibnamefont {Slama}},\ and\ \bibinfo {author} {\bibfnamefont {C.}~\bibnamefont {Zimmermann}},\ }\href {https://doi.org/10.1103/PhysRevLett.112.115302} {\bibfield  {journal} {\bibinfo  {journal} {Phys. Rev. Lett.}\ }\textbf {\bibinfo {volume} {112}},\ \bibinfo {pages} {115302} (\bibinfo {year} {2014})}\BibitemShut {NoStop}%
\bibitem [{\citenamefont {Gangl}\ and\ \citenamefont {Ritsch}(2000)}]{Gangl2000}%
  \BibitemOpen
  \bibfield  {author} {\bibinfo {author} {\bibfnamefont {M.}~\bibnamefont {Gangl}}\ and\ \bibinfo {author} {\bibfnamefont {H.}~\bibnamefont {Ritsch}},\ }\href {https://doi.org/10.1103/PhysRevA.61.043405} {\bibfield  {journal} {\bibinfo  {journal} {Phys. Rev. A}\ }\textbf {\bibinfo {volume} {61}},\ \bibinfo {pages} {043405} (\bibinfo {year} {2000})}\BibitemShut {NoStop}%
\bibitem [{\citenamefont {Mivehvar}\ \emph {et~al.}(2018)\citenamefont {Mivehvar}, \citenamefont {Ostermann}, \citenamefont {Piazza},\ and\ \citenamefont {Ritsch}}]{Mivehvar2018}%
  \BibitemOpen
  \bibfield  {author} {\bibinfo {author} {\bibfnamefont {F.}~\bibnamefont {Mivehvar}}, \bibinfo {author} {\bibfnamefont {S.}~\bibnamefont {Ostermann}}, \bibinfo {author} {\bibfnamefont {F.}~\bibnamefont {Piazza}},\ and\ \bibinfo {author} {\bibfnamefont {H.}~\bibnamefont {Ritsch}},\ }\href {https://doi.org/10.1103/PhysRevLett.120.123601} {\bibfield  {journal} {\bibinfo  {journal} {Phys. Rev. Lett.}\ }\textbf {\bibinfo {volume} {120}},\ \bibinfo {pages} {123601} (\bibinfo {year} {2018})}\BibitemShut {NoStop}%
\bibitem [{\citenamefont {Klinner}\ \emph {et~al.}(2006)\citenamefont {Klinner}, \citenamefont {Lindholdt}, \citenamefont {Nagorny},\ and\ \citenamefont {Hemmerich}}]{Klinner2006}%
  \BibitemOpen
  \bibfield  {author} {\bibinfo {author} {\bibfnamefont {J.}~\bibnamefont {Klinner}}, \bibinfo {author} {\bibfnamefont {M.}~\bibnamefont {Lindholdt}}, \bibinfo {author} {\bibfnamefont {B.}~\bibnamefont {Nagorny}},\ and\ \bibinfo {author} {\bibfnamefont {A.}~\bibnamefont {Hemmerich}},\ }\href {https://doi.org/10.1103/PhysRevLett.96.023002} {\bibfield  {journal} {\bibinfo  {journal} {Phys. Rev. Lett.}\ }\textbf {\bibinfo {volume} {96}},\ \bibinfo {pages} {023002} (\bibinfo {year} {2006})}\BibitemShut {NoStop}%
\bibitem [{\citenamefont {Kollár}\ \emph {et~al.}(2015)\citenamefont {Kollár}, \citenamefont {Papageorge}, \citenamefont {Baumann}, \citenamefont {Armen},\ and\ \citenamefont {Lev}}]{Kollár2015}%
  \BibitemOpen
  \bibfield  {author} {\bibinfo {author} {\bibfnamefont {A.~J.}\ \bibnamefont {Kollár}}, \bibinfo {author} {\bibfnamefont {A.~T.}\ \bibnamefont {Papageorge}}, \bibinfo {author} {\bibfnamefont {K.}~\bibnamefont {Baumann}}, \bibinfo {author} {\bibfnamefont {M.~A.}\ \bibnamefont {Armen}},\ and\ \bibinfo {author} {\bibfnamefont {B.~L.}\ \bibnamefont {Lev}},\ }\href {https://doi.org/10.1088/1367-2630/17/4/043012} {\bibfield  {journal} {\bibinfo  {journal} {New Journal of Physics}\ }\textbf {\bibinfo {volume} {17}},\ \bibinfo {pages} {043012} (\bibinfo {year} {2015})}\BibitemShut {NoStop}%
\bibitem [{\citenamefont {Vaidya}\ \emph {et~al.}(2018)\citenamefont {Vaidya}, \citenamefont {Guo}, \citenamefont {Kroeze}, \citenamefont {Ballantine}, \citenamefont {Koll\'ar}, \citenamefont {Keeling},\ and\ \citenamefont {Lev}}]{Vaidya2018}%
  \BibitemOpen
  \bibfield  {author} {\bibinfo {author} {\bibfnamefont {V.~D.}\ \bibnamefont {Vaidya}}, \bibinfo {author} {\bibfnamefont {Y.}~\bibnamefont {Guo}}, \bibinfo {author} {\bibfnamefont {R.~M.}\ \bibnamefont {Kroeze}}, \bibinfo {author} {\bibfnamefont {K.~E.}\ \bibnamefont {Ballantine}}, \bibinfo {author} {\bibfnamefont {A.~J.}\ \bibnamefont {Koll\'ar}}, \bibinfo {author} {\bibfnamefont {J.}~\bibnamefont {Keeling}},\ and\ \bibinfo {author} {\bibfnamefont {B.~L.}\ \bibnamefont {Lev}},\ }\href {https://doi.org/10.1103/PhysRevX.8.011002} {\bibfield  {journal} {\bibinfo  {journal} {Phys. Rev. X}\ }\textbf {\bibinfo {volume} {8}},\ \bibinfo {pages} {011002} (\bibinfo {year} {2018})}\BibitemShut {NoStop}%
\bibitem [{\citenamefont {Gopalakrishnan}\ \emph {et~al.}(2011)\citenamefont {Gopalakrishnan}, \citenamefont {Lev},\ and\ \citenamefont {Goldbart}}]{Gopalakrishnan11}%
  \BibitemOpen
  \bibfield  {author} {\bibinfo {author} {\bibfnamefont {S.}~\bibnamefont {Gopalakrishnan}}, \bibinfo {author} {\bibfnamefont {B.~L.}\ \bibnamefont {Lev}},\ and\ \bibinfo {author} {\bibfnamefont {P.~M.}\ \bibnamefont {Goldbart}},\ }\href {https://doi.org/10.1103/PhysRevLett.107.277201} {\bibfield  {journal} {\bibinfo  {journal} {Phys. Rev. Lett.}\ }\textbf {\bibinfo {volume} {107}},\ \bibinfo {pages} {277201} (\bibinfo {year} {2011})}\BibitemShut {NoStop}%
\bibitem [{\citenamefont {Gopalakrishnan}\ \emph {et~al.}(2009)\citenamefont {Gopalakrishnan}, \citenamefont {Lev},\ and\ \citenamefont {Goldbart}}]{Gopalakrishnan2009}%
  \BibitemOpen
  \bibfield  {author} {\bibinfo {author} {\bibfnamefont {S.}~\bibnamefont {Gopalakrishnan}}, \bibinfo {author} {\bibfnamefont {B.~L.}\ \bibnamefont {Lev}},\ and\ \bibinfo {author} {\bibfnamefont {P.~M.}\ \bibnamefont {Goldbart}},\ }\href {https://doi.org/10.1038/nphys1403} {\bibfield  {journal} {\bibinfo  {journal} {Nature Physics}\ }\textbf {\bibinfo {volume} {5}},\ \bibinfo {pages} {845} (\bibinfo {year} {2009})}\BibitemShut {NoStop}%
\bibitem [{\citenamefont {Ballantine}\ \emph {et~al.}(2017)\citenamefont {Ballantine}, \citenamefont {Lev},\ and\ \citenamefont {Keeling}}]{Ballantine17}%
  \BibitemOpen
  \bibfield  {author} {\bibinfo {author} {\bibfnamefont {K.~E.}\ \bibnamefont {Ballantine}}, \bibinfo {author} {\bibfnamefont {B.~L.}\ \bibnamefont {Lev}},\ and\ \bibinfo {author} {\bibfnamefont {J.}~\bibnamefont {Keeling}},\ }\href {https://doi.org/10.1103/PhysRevLett.118.045302} {\bibfield  {journal} {\bibinfo  {journal} {Phys. Rev. Lett.}\ }\textbf {\bibinfo {volume} {118}},\ \bibinfo {pages} {045302} (\bibinfo {year} {2017})}\BibitemShut {NoStop}%
\bibitem [{\citenamefont {Torggler}\ \emph {et~al.}(2017)\citenamefont {Torggler}, \citenamefont {Kr\"amer},\ and\ \citenamefont {Ritsch}}]{Torggler17}%
  \BibitemOpen
  \bibfield  {author} {\bibinfo {author} {\bibfnamefont {V.}~\bibnamefont {Torggler}}, \bibinfo {author} {\bibfnamefont {S.}~\bibnamefont {Kr\"amer}},\ and\ \bibinfo {author} {\bibfnamefont {H.}~\bibnamefont {Ritsch}},\ }\href {https://doi.org/10.1103/PhysRevA.95.032310} {\bibfield  {journal} {\bibinfo  {journal} {Phys. Rev. A}\ }\textbf {\bibinfo {volume} {95}},\ \bibinfo {pages} {032310} (\bibinfo {year} {2017})}\BibitemShut {NoStop}%
\bibitem [{\citenamefont {Rotondo}\ \emph {et~al.}(2018)\citenamefont {Rotondo}, \citenamefont {Marcuzzi}, \citenamefont {Garrahan}, \citenamefont {Lesanovsky},\ and\ \citenamefont {Müller}}]{Rotondo2018}%
  \BibitemOpen
  \bibfield  {author} {\bibinfo {author} {\bibfnamefont {P.}~\bibnamefont {Rotondo}}, \bibinfo {author} {\bibfnamefont {M.}~\bibnamefont {Marcuzzi}}, \bibinfo {author} {\bibfnamefont {J.~P.}\ \bibnamefont {Garrahan}}, \bibinfo {author} {\bibfnamefont {I.}~\bibnamefont {Lesanovsky}},\ and\ \bibinfo {author} {\bibfnamefont {M.}~\bibnamefont {Müller}},\ }\href {https://doi.org/10.1088/1751-8121/aaabcb} {\bibfield  {journal} {\bibinfo  {journal} {Journal of Physics A: Mathematical and Theoretical}\ }\textbf {\bibinfo {volume} {51}},\ \bibinfo {pages} {115301} (\bibinfo {year} {2018})}\BibitemShut {NoStop}%
\bibitem [{\citenamefont {Fiorelli}\ \emph {et~al.}(2020)\citenamefont {Fiorelli}, \citenamefont {Marcuzzi}, \citenamefont {Rotondo}, \citenamefont {Carollo},\ and\ \citenamefont {Lesanovsky}}]{Fiorelli2020}%
  \BibitemOpen
  \bibfield  {author} {\bibinfo {author} {\bibfnamefont {E.}~\bibnamefont {Fiorelli}}, \bibinfo {author} {\bibfnamefont {M.}~\bibnamefont {Marcuzzi}}, \bibinfo {author} {\bibfnamefont {P.}~\bibnamefont {Rotondo}}, \bibinfo {author} {\bibfnamefont {F.}~\bibnamefont {Carollo}},\ and\ \bibinfo {author} {\bibfnamefont {I.}~\bibnamefont {Lesanovsky}},\ }\href {https://doi.org/10.1103/PhysRevLett.125.070604} {\bibfield  {journal} {\bibinfo  {journal} {Phys. Rev. Lett.}\ }\textbf {\bibinfo {volume} {125}},\ \bibinfo {pages} {070604} (\bibinfo {year} {2020})}\BibitemShut {NoStop}%
\bibitem [{\citenamefont {Marsh}\ \emph {et~al.}(2021)\citenamefont {Marsh}, \citenamefont {Guo}, \citenamefont {Kroeze}, \citenamefont {Gopalakrishnan}, \citenamefont {Ganguli}, \citenamefont {Keeling},\ and\ \citenamefont {Lev}}]{Marsh2021}%
  \BibitemOpen
  \bibfield  {author} {\bibinfo {author} {\bibfnamefont {B.~P.}\ \bibnamefont {Marsh}}, \bibinfo {author} {\bibfnamefont {Y.}~\bibnamefont {Guo}}, \bibinfo {author} {\bibfnamefont {R.~M.}\ \bibnamefont {Kroeze}}, \bibinfo {author} {\bibfnamefont {S.}~\bibnamefont {Gopalakrishnan}}, \bibinfo {author} {\bibfnamefont {S.}~\bibnamefont {Ganguli}}, \bibinfo {author} {\bibfnamefont {J.}~\bibnamefont {Keeling}},\ and\ \bibinfo {author} {\bibfnamefont {B.~L.}\ \bibnamefont {Lev}},\ }\href {https://doi.org/10.1103/PhysRevX.11.021048} {\bibfield  {journal} {\bibinfo  {journal} {Phys. Rev. X}\ }\textbf {\bibinfo {volume} {11}},\ \bibinfo {pages} {021048} (\bibinfo {year} {2021})}\BibitemShut {NoStop}%
\bibitem [{\citenamefont {Bentsen}\ \emph {et~al.}(2019)\citenamefont {Bentsen}, \citenamefont {Hashizume}, \citenamefont {Buyskikh}, \citenamefont {Davis}, \citenamefont {Daley}, \citenamefont {Gubser},\ and\ \citenamefont {Schleier-Smith}}]{Bentsen2019}%
  \BibitemOpen
  \bibfield  {author} {\bibinfo {author} {\bibfnamefont {G.}~\bibnamefont {Bentsen}}, \bibinfo {author} {\bibfnamefont {T.}~\bibnamefont {Hashizume}}, \bibinfo {author} {\bibfnamefont {A.~S.}\ \bibnamefont {Buyskikh}}, \bibinfo {author} {\bibfnamefont {E.~J.}\ \bibnamefont {Davis}}, \bibinfo {author} {\bibfnamefont {A.~J.}\ \bibnamefont {Daley}}, \bibinfo {author} {\bibfnamefont {S.~S.}\ \bibnamefont {Gubser}},\ and\ \bibinfo {author} {\bibfnamefont {M.}~\bibnamefont {Schleier-Smith}},\ }\href {https://doi.org/10.1103/PhysRevLett.123.130601} {\bibfield  {journal} {\bibinfo  {journal} {Phys. Rev. Lett.}\ }\textbf {\bibinfo {volume} {123}},\ \bibinfo {pages} {130601} (\bibinfo {year} {2019})}\BibitemShut {NoStop}%
\bibitem [{\citenamefont {Baumann}\ \emph {et~al.}(2010)\citenamefont {Baumann}, \citenamefont {Guerlin}, \citenamefont {Brennecke},\ and\ \citenamefont {Esslinger}}]{Baumann2010}%
  \BibitemOpen
  \bibfield  {author} {\bibinfo {author} {\bibfnamefont {K.}~\bibnamefont {Baumann}}, \bibinfo {author} {\bibfnamefont {C.}~\bibnamefont {Guerlin}}, \bibinfo {author} {\bibfnamefont {F.}~\bibnamefont {Brennecke}},\ and\ \bibinfo {author} {\bibfnamefont {T.}~\bibnamefont {Esslinger}},\ }\href {https://doi.org/10.1038/nature09009} {\bibfield  {journal} {\bibinfo  {journal} {Nature}\ }\textbf {\bibinfo {volume} {464}},\ \bibinfo {pages} {1301} (\bibinfo {year} {2010})}\BibitemShut {NoStop}%
\bibitem [{\citenamefont {Klinder}\ \emph {et~al.}(2015)\citenamefont {Klinder}, \citenamefont {Keßler}, \citenamefont {Wolke}, \citenamefont {Mathey},\ and\ \citenamefont {Hemmerich}}]{Klinder15}%
  \BibitemOpen
  \bibfield  {author} {\bibinfo {author} {\bibfnamefont {J.}~\bibnamefont {Klinder}}, \bibinfo {author} {\bibfnamefont {H.}~\bibnamefont {Keßler}}, \bibinfo {author} {\bibfnamefont {M.}~\bibnamefont {Wolke}}, \bibinfo {author} {\bibfnamefont {L.}~\bibnamefont {Mathey}},\ and\ \bibinfo {author} {\bibfnamefont {A.}~\bibnamefont {Hemmerich}},\ }\href {https://doi.org/10.1073/pnas.1417132112} {\bibfield  {journal} {\bibinfo  {journal} {Proceedings of the National Academy of Sciences}\ }\textbf {\bibinfo {volume} {112}},\ \bibinfo {pages} {3290} (\bibinfo {year} {2015})},\ \Eprint {https://arxiv.org/abs/https://www.pnas.org/doi/pdf/10.1073/pnas.1417132112} {https://www.pnas.org/doi/pdf/10.1073/pnas.1417132112} \BibitemShut {NoStop}%
\bibitem [{\citenamefont {Benary}\ \emph {et~al.}(2022)\citenamefont {Benary}, \citenamefont {Baals}, \citenamefont {Bernhart}, \citenamefont {Jiang}, \citenamefont {Röhrle},\ and\ \citenamefont {Ott}}]{Benary2022}%
  \BibitemOpen
  \bibfield  {author} {\bibinfo {author} {\bibfnamefont {J.}~\bibnamefont {Benary}}, \bibinfo {author} {\bibfnamefont {C.}~\bibnamefont {Baals}}, \bibinfo {author} {\bibfnamefont {E.}~\bibnamefont {Bernhart}}, \bibinfo {author} {\bibfnamefont {J.}~\bibnamefont {Jiang}}, \bibinfo {author} {\bibfnamefont {M.}~\bibnamefont {Röhrle}},\ and\ \bibinfo {author} {\bibfnamefont {H.}~\bibnamefont {Ott}},\ }\href {https://doi.org/10.1088/1367-2630/ac97b6} {\bibfield  {journal} {\bibinfo  {journal} {New Journal of Physics}\ }\textbf {\bibinfo {volume} {24}},\ \bibinfo {pages} {103034} (\bibinfo {year} {2022})}\BibitemShut {NoStop}%
\bibitem [{\citenamefont {Ferri}\ \emph {et~al.}(2021)\citenamefont {Ferri}, \citenamefont {Rosa-Medina}, \citenamefont {Finger}, \citenamefont {Dogra}, \citenamefont {Soriente}, \citenamefont {Zilberberg}, \citenamefont {Donner},\ and\ \citenamefont {Esslinger}}]{Ferri2021}%
  \BibitemOpen
  \bibfield  {author} {\bibinfo {author} {\bibfnamefont {F.}~\bibnamefont {Ferri}}, \bibinfo {author} {\bibfnamefont {R.}~\bibnamefont {Rosa-Medina}}, \bibinfo {author} {\bibfnamefont {F.}~\bibnamefont {Finger}}, \bibinfo {author} {\bibfnamefont {N.}~\bibnamefont {Dogra}}, \bibinfo {author} {\bibfnamefont {M.}~\bibnamefont {Soriente}}, \bibinfo {author} {\bibfnamefont {O.}~\bibnamefont {Zilberberg}}, \bibinfo {author} {\bibfnamefont {T.}~\bibnamefont {Donner}},\ and\ \bibinfo {author} {\bibfnamefont {T.}~\bibnamefont {Esslinger}},\ }\href {https://doi.org/10.1103/PhysRevX.11.041046} {\bibfield  {journal} {\bibinfo  {journal} {Phys. Rev. X}\ }\textbf {\bibinfo {volume} {11}},\ \bibinfo {pages} {041046} (\bibinfo {year} {2021})}\BibitemShut {NoStop}%
\bibitem [{\citenamefont {Greiner}\ \emph {et~al.}(2002)\citenamefont {Greiner}, \citenamefont {Mandel}, \citenamefont {Esslinger}, \citenamefont {H{\"a}nsch},\ and\ \citenamefont {Bloch}}]{Greiner2002}%
  \BibitemOpen
  \bibfield  {author} {\bibinfo {author} {\bibfnamefont {M.}~\bibnamefont {Greiner}}, \bibinfo {author} {\bibfnamefont {O.}~\bibnamefont {Mandel}}, \bibinfo {author} {\bibfnamefont {T.}~\bibnamefont {Esslinger}}, \bibinfo {author} {\bibfnamefont {T.~W.}\ \bibnamefont {H{\"a}nsch}},\ and\ \bibinfo {author} {\bibfnamefont {I.}~\bibnamefont {Bloch}},\ }\href {https://doi.org/10.1038/415039a} {\bibfield  {journal} {\bibinfo  {journal} {Nature}\ }\textbf {\bibinfo {volume} {415}},\ \bibinfo {pages} {39} (\bibinfo {year} {2002})}\BibitemShut {NoStop}%
\bibitem [{\citenamefont {Chen}\ \emph {et~al.}(2005)\citenamefont {Chen}, \citenamefont {Stajic}, \citenamefont {Tan},\ and\ \citenamefont {Levin}}]{Chen2005}%
  \BibitemOpen
  \bibfield  {author} {\bibinfo {author} {\bibfnamefont {Q.}~\bibnamefont {Chen}}, \bibinfo {author} {\bibfnamefont {J.}~\bibnamefont {Stajic}}, \bibinfo {author} {\bibfnamefont {S.}~\bibnamefont {Tan}},\ and\ \bibinfo {author} {\bibfnamefont {K.}~\bibnamefont {Levin}},\ }\href {https://doi.org/https://doi.org/10.1016/j.physrep.2005.02.005} {\bibfield  {journal} {\bibinfo  {journal} {Physics Reports}\ }\textbf {\bibinfo {volume} {412}},\ \bibinfo {pages} {1} (\bibinfo {year} {2005})}\BibitemShut {NoStop}%
\bibitem [{\citenamefont {Jin}\ \emph {et~al.}(2016)\citenamefont {Jin}, \citenamefont {Biella}, \citenamefont {Viyuela}, \citenamefont {Mazza}, \citenamefont {Keeling}, \citenamefont {Fazio},\ and\ \citenamefont {Rossini}}]{Jin2016}%
  \BibitemOpen
  \bibfield  {author} {\bibinfo {author} {\bibfnamefont {J.}~\bibnamefont {Jin}}, \bibinfo {author} {\bibfnamefont {A.}~\bibnamefont {Biella}}, \bibinfo {author} {\bibfnamefont {O.}~\bibnamefont {Viyuela}}, \bibinfo {author} {\bibfnamefont {L.}~\bibnamefont {Mazza}}, \bibinfo {author} {\bibfnamefont {J.}~\bibnamefont {Keeling}}, \bibinfo {author} {\bibfnamefont {R.}~\bibnamefont {Fazio}},\ and\ \bibinfo {author} {\bibfnamefont {D.}~\bibnamefont {Rossini}},\ }\href {https://doi.org/10.1103/PhysRevX.6.031011} {\bibfield  {journal} {\bibinfo  {journal} {Phys. Rev. X}\ }\textbf {\bibinfo {volume} {6}},\ \bibinfo {pages} {031011} (\bibinfo {year} {2016})}\BibitemShut {NoStop}%
\bibitem [{\citenamefont {Lee}\ \emph {et~al.}(2013)\citenamefont {Lee}, \citenamefont {Gopalakrishnan},\ and\ \citenamefont {Lukin}}]{Lee2013}%
  \BibitemOpen
  \bibfield  {author} {\bibinfo {author} {\bibfnamefont {T.~E.}\ \bibnamefont {Lee}}, \bibinfo {author} {\bibfnamefont {S.}~\bibnamefont {Gopalakrishnan}},\ and\ \bibinfo {author} {\bibfnamefont {M.~D.}\ \bibnamefont {Lukin}},\ }\href {https://doi.org/10.1103/PhysRevLett.110.257204} {\bibfield  {journal} {\bibinfo  {journal} {Phys. Rev. Lett.}\ }\textbf {\bibinfo {volume} {110}},\ \bibinfo {pages} {257204} (\bibinfo {year} {2013})}\BibitemShut {NoStop}%
\bibitem [{\citenamefont {Tindall}\ \emph {et~al.}(2019)\citenamefont {Tindall}, \citenamefont {Bu\ifmmode~\check{c}\else \v{c}\fi{}a}, \citenamefont {Coulthard},\ and\ \citenamefont {Jaksch}}]{tindall}%
  \BibitemOpen
  \bibfield  {author} {\bibinfo {author} {\bibfnamefont {J.}~\bibnamefont {Tindall}}, \bibinfo {author} {\bibfnamefont {B.}~\bibnamefont {Bu\ifmmode~\check{c}\else \v{c}\fi{}a}}, \bibinfo {author} {\bibfnamefont {J.~R.}\ \bibnamefont {Coulthard}},\ and\ \bibinfo {author} {\bibfnamefont {D.}~\bibnamefont {Jaksch}},\ }\bibfield  {journal} {\bibinfo  {journal} {Phys. Rev. Lett.}\ }\href {https://doi.org/10.1103/PhysRevLett.123.030603} {10.1103/PhysRevLett.123.030603} (\bibinfo {year} {2019})\BibitemShut {NoStop}%
\bibitem [{\citenamefont {Lindblad}(1976)}]{Lindblad1976}%
  \BibitemOpen
  \bibfield  {author} {\bibinfo {author} {\bibfnamefont {G.}~\bibnamefont {Lindblad}},\ }\href {https://doi.org/10.1007/BF01608499} {\bibfield  {journal} {\bibinfo  {journal} {Communications in Mathematical Physics}\ }\textbf {\bibinfo {volume} {48}},\ \bibinfo {pages} {119} (\bibinfo {year} {1976})}\BibitemShut {NoStop}%
\bibitem [{\citenamefont {Gorini}\ \emph {et~al.}(1976)\citenamefont {Gorini}, \citenamefont {Kossakowski},\ and\ \citenamefont {Sudarshan}}]{Gorini1976}%
  \BibitemOpen
  \bibfield  {author} {\bibinfo {author} {\bibfnamefont {V.}~\bibnamefont {Gorini}}, \bibinfo {author} {\bibfnamefont {A.}~\bibnamefont {Kossakowski}},\ and\ \bibinfo {author} {\bibfnamefont {E.~C.~G.}\ \bibnamefont {Sudarshan}},\ }\href {https://doi.org/10.1063/1.522979} {\bibfield  {journal} {\bibinfo  {journal} {Journal of Mathematical Physics}\ }\textbf {\bibinfo {volume} {17}},\ \bibinfo {pages} {821} (\bibinfo {year} {1976})},\ \Eprint {https://arxiv.org/abs/https://pubs.aip.org/aip/jmp/article-pdf/17/5/821/19090720/821\_1\_online.pdf} {https://pubs.aip.org/aip/jmp/article-pdf/17/5/821/19090720/821\_1\_online.pdf} \BibitemShut {NoStop}%
\bibitem [{\citenamefont {Yang}(1989)}]{Yang1989}%
  \BibitemOpen
  \bibfield  {author} {\bibinfo {author} {\bibfnamefont {C.~N.}\ \bibnamefont {Yang}},\ }\href {https://doi.org/10.1103/PhysRevLett.63.2144} {\bibfield  {journal} {\bibinfo  {journal} {Phys. Rev. Lett.}\ }\textbf {\bibinfo {volume} {63}},\ \bibinfo {pages} {2144} (\bibinfo {year} {1989})}\BibitemShut {NoStop}%
\bibitem [{\citenamefont {Agterberg}\ \emph {et~al.}(2020)\citenamefont {Agterberg}, \citenamefont {Davis}, \citenamefont {Edkins}, \citenamefont {Fradkin}, \citenamefont {Van~Harlingen}, \citenamefont {Kivelson}, \citenamefont {Lee}, \citenamefont {Radzihovsky}, \citenamefont {Tranquada},\ and\ \citenamefont {Wang}}]{Agterberg_2020}%
  \BibitemOpen
  \bibfield  {author} {\bibinfo {author} {\bibfnamefont {D.~F.}\ \bibnamefont {Agterberg}}, \bibinfo {author} {\bibfnamefont {J.~S.}\ \bibnamefont {Davis}}, \bibinfo {author} {\bibfnamefont {S.~D.}\ \bibnamefont {Edkins}}, \bibinfo {author} {\bibfnamefont {E.}~\bibnamefont {Fradkin}}, \bibinfo {author} {\bibfnamefont {D.~J.}\ \bibnamefont {Van~Harlingen}}, \bibinfo {author} {\bibfnamefont {S.~A.}\ \bibnamefont {Kivelson}}, \bibinfo {author} {\bibfnamefont {P.~A.}\ \bibnamefont {Lee}}, \bibinfo {author} {\bibfnamefont {L.}~\bibnamefont {Radzihovsky}}, \bibinfo {author} {\bibfnamefont {J.~M.}\ \bibnamefont {Tranquada}},\ and\ \bibinfo {author} {\bibfnamefont {Y.}~\bibnamefont {Wang}},\ }\href {https://doi.org/10.1146/annurev-conmatphys-031119-050711} {\bibfield  {journal} {\bibinfo  {journal} {Annual Review of Condensed Matter Physics}\ }\textbf {\bibinfo {volume} {11}},\ \bibinfo {pages} {231–270} (\bibinfo {year} {2020})}\BibitemShut {NoStop}%
\bibitem [{\citenamefont {Zhao}\ \emph {et~al.}(2023)\citenamefont {Zhao}, \citenamefont {Blackwell}, \citenamefont {Thinel}, \citenamefont {Handa}, \citenamefont {Ishida}, \citenamefont {Zhu}, \citenamefont {Iyo}, \citenamefont {Eisaki}, \citenamefont {Pasupathy},\ and\ \citenamefont {Fujita}}]{Zhao_2023}%
  \BibitemOpen
  \bibfield  {author} {\bibinfo {author} {\bibfnamefont {H.}~\bibnamefont {Zhao}}, \bibinfo {author} {\bibfnamefont {R.}~\bibnamefont {Blackwell}}, \bibinfo {author} {\bibfnamefont {M.}~\bibnamefont {Thinel}}, \bibinfo {author} {\bibfnamefont {T.}~\bibnamefont {Handa}}, \bibinfo {author} {\bibfnamefont {S.}~\bibnamefont {Ishida}}, \bibinfo {author} {\bibfnamefont {X.}~\bibnamefont {Zhu}}, \bibinfo {author} {\bibfnamefont {A.}~\bibnamefont {Iyo}}, \bibinfo {author} {\bibfnamefont {H.}~\bibnamefont {Eisaki}}, \bibinfo {author} {\bibfnamefont {A.~N.}\ \bibnamefont {Pasupathy}},\ and\ \bibinfo {author} {\bibfnamefont {K.}~\bibnamefont {Fujita}},\ }\href {https://doi.org/10.1038/s41586-023-06103-7} {\bibfield  {journal} {\bibinfo  {journal} {Nature}\ }\textbf {\bibinfo {volume} {618}},\ \bibinfo {pages} {940–945} (\bibinfo {year} {2023})}\BibitemShut {NoStop}%
\bibitem [{\citenamefont {Zhang}\ \emph {et~al.}(2024)\citenamefont {Zhang}, \citenamefont {Hu},\ and\ \citenamefont {Neupert}}]{Zhang_2024}%
  \BibitemOpen
  \bibfield  {author} {\bibinfo {author} {\bibfnamefont {S.-B.}\ \bibnamefont {Zhang}}, \bibinfo {author} {\bibfnamefont {L.-H.}\ \bibnamefont {Hu}},\ and\ \bibinfo {author} {\bibfnamefont {T.}~\bibnamefont {Neupert}},\ }\bibfield  {journal} {\bibinfo  {journal} {Nature Communications}\ }\textbf {\bibinfo {volume} {15}},\ \href {https://doi.org/10.1038/s41467-024-45951-3} {10.1038/s41467-024-45951-3} (\bibinfo {year} {2024})\BibitemShut {NoStop}%
\bibitem [{\citenamefont {Colella}\ \emph {et~al.}(2018)\citenamefont {Colella}, \citenamefont {Citro}, \citenamefont {Barsanti}, \citenamefont {Rossini},\ and\ \citenamefont {Chiofalo}}]{Colella2018}%
  \BibitemOpen
  \bibfield  {author} {\bibinfo {author} {\bibfnamefont {E.}~\bibnamefont {Colella}}, \bibinfo {author} {\bibfnamefont {R.}~\bibnamefont {Citro}}, \bibinfo {author} {\bibfnamefont {M.}~\bibnamefont {Barsanti}}, \bibinfo {author} {\bibfnamefont {D.}~\bibnamefont {Rossini}},\ and\ \bibinfo {author} {\bibfnamefont {M.-L.}\ \bibnamefont {Chiofalo}},\ }\href {https://doi.org/10.1103/PhysRevB.97.134502} {\bibfield  {journal} {\bibinfo  {journal} {Phys. Rev. B}\ }\textbf {\bibinfo {volume} {97}},\ \bibinfo {pages} {134502} (\bibinfo {year} {2018})}\BibitemShut {NoStop}%
\bibitem [{\citenamefont {Schlawin}\ and\ \citenamefont {Jaksch}(2019)}]{Schlawin2019}%
  \BibitemOpen
  \bibfield  {author} {\bibinfo {author} {\bibfnamefont {F.}~\bibnamefont {Schlawin}}\ and\ \bibinfo {author} {\bibfnamefont {D.}~\bibnamefont {Jaksch}},\ }\href {https://doi.org/10.1103/PhysRevLett.123.133601} {\bibfield  {journal} {\bibinfo  {journal} {Phys. Rev. Lett.}\ }\textbf {\bibinfo {volume} {123}},\ \bibinfo {pages} {133601} (\bibinfo {year} {2019})}\BibitemShut {NoStop}%
\bibitem [{\citenamefont {Camacho-Guardian}\ \emph {et~al.}(2017)\citenamefont {Camacho-Guardian}, \citenamefont {Paredes},\ and\ \citenamefont {Caballero-Ben\'{\i}tez}}]{Camacho2017}%
  \BibitemOpen
  \bibfield  {author} {\bibinfo {author} {\bibfnamefont {A.}~\bibnamefont {Camacho-Guardian}}, \bibinfo {author} {\bibfnamefont {R.}~\bibnamefont {Paredes}},\ and\ \bibinfo {author} {\bibfnamefont {S.~F.}\ \bibnamefont {Caballero-Ben\'{\i}tez}},\ }\href {https://doi.org/10.1103/PhysRevA.96.051602} {\bibfield  {journal} {\bibinfo  {journal} {Phys. Rev. A}\ }\textbf {\bibinfo {volume} {96}},\ \bibinfo {pages} {051602} (\bibinfo {year} {2017})}\BibitemShut {NoStop}%
\bibitem [{\citenamefont {Davis}\ \emph {et~al.}(2019)\citenamefont {Davis}, \citenamefont {Bentsen}, \citenamefont {Homeier}, \citenamefont {Li},\ and\ \citenamefont {Schleier-Smith}}]{Davis2019}%
  \BibitemOpen
  \bibfield  {author} {\bibinfo {author} {\bibfnamefont {E.~J.}\ \bibnamefont {Davis}}, \bibinfo {author} {\bibfnamefont {G.}~\bibnamefont {Bentsen}}, \bibinfo {author} {\bibfnamefont {L.}~\bibnamefont {Homeier}}, \bibinfo {author} {\bibfnamefont {T.}~\bibnamefont {Li}},\ and\ \bibinfo {author} {\bibfnamefont {M.~H.}\ \bibnamefont {Schleier-Smith}},\ }\href {https://doi.org/10.1103/PhysRevLett.122.010405} {\bibfield  {journal} {\bibinfo  {journal} {Phys. Rev. Lett.}\ }\textbf {\bibinfo {volume} {122}},\ \bibinfo {pages} {010405} (\bibinfo {year} {2019})}\BibitemShut {NoStop}%
\bibitem [{\citenamefont {Norcia}\ \emph {et~al.}(2018)\citenamefont {Norcia}, \citenamefont {Lewis-Swan}, \citenamefont {Cline}, \citenamefont {Zhu}, \citenamefont {Rey},\ and\ \citenamefont {Thompson}}]{Norcia2018}%
  \BibitemOpen
  \bibfield  {author} {\bibinfo {author} {\bibfnamefont {M.~A.}\ \bibnamefont {Norcia}}, \bibinfo {author} {\bibfnamefont {R.~J.}\ \bibnamefont {Lewis-Swan}}, \bibinfo {author} {\bibfnamefont {J.~R.~K.}\ \bibnamefont {Cline}}, \bibinfo {author} {\bibfnamefont {B.}~\bibnamefont {Zhu}}, \bibinfo {author} {\bibfnamefont {A.~M.}\ \bibnamefont {Rey}},\ and\ \bibinfo {author} {\bibfnamefont {J.~K.}\ \bibnamefont {Thompson}},\ }\href {https://doi.org/10.1126/science.aar3102} {\bibfield  {journal} {\bibinfo  {journal} {Science}\ }\textbf {\bibinfo {volume} {361}},\ \bibinfo {pages} {259} (\bibinfo {year} {2018})},\ \Eprint {https://arxiv.org/abs/https://www.science.org/doi/pdf/10.1126/science.aar3102} {https://www.science.org/doi/pdf/10.1126/science.aar3102} \BibitemShut {NoStop}%
\bibitem [{\citenamefont {Baier}\ \emph {et~al.}(2018)\citenamefont {Baier}, \citenamefont {Petter}, \citenamefont {Becher}, \citenamefont {Patscheider}, \citenamefont {Natale}, \citenamefont {Chomaz}, \citenamefont {Mark},\ and\ \citenamefont {Ferlaino}}]{Baier2018}%
  \BibitemOpen
  \bibfield  {author} {\bibinfo {author} {\bibfnamefont {S.}~\bibnamefont {Baier}}, \bibinfo {author} {\bibfnamefont {D.}~\bibnamefont {Petter}}, \bibinfo {author} {\bibfnamefont {J.~H.}\ \bibnamefont {Becher}}, \bibinfo {author} {\bibfnamefont {A.}~\bibnamefont {Patscheider}}, \bibinfo {author} {\bibfnamefont {G.}~\bibnamefont {Natale}}, \bibinfo {author} {\bibfnamefont {L.}~\bibnamefont {Chomaz}}, \bibinfo {author} {\bibfnamefont {M.~J.}\ \bibnamefont {Mark}},\ and\ \bibinfo {author} {\bibfnamefont {F.}~\bibnamefont {Ferlaino}},\ }\href {https://doi.org/10.1103/PhysRevLett.121.093602} {\bibfield  {journal} {\bibinfo  {journal} {Phys. Rev. Lett.}\ }\textbf {\bibinfo {volume} {121}},\ \bibinfo {pages} {093602} (\bibinfo {year} {2018})}\BibitemShut {NoStop}%
\bibitem [{\citenamefont {Mazurenko}\ \emph {et~al.}(2017)\citenamefont {Mazurenko}, \citenamefont {Chiu}, \citenamefont {Ji}, \citenamefont {Parsons}, \citenamefont {Kan{\'a}sz-Nagy}, \citenamefont {Schmidt}, \citenamefont {Grusdt}, \citenamefont {Demler}, \citenamefont {Greif},\ and\ \citenamefont {Greiner}}]{Mazurenko2017}%
  \BibitemOpen
  \bibfield  {author} {\bibinfo {author} {\bibfnamefont {A.}~\bibnamefont {Mazurenko}}, \bibinfo {author} {\bibfnamefont {C.~S.}\ \bibnamefont {Chiu}}, \bibinfo {author} {\bibfnamefont {G.}~\bibnamefont {Ji}}, \bibinfo {author} {\bibfnamefont {M.~F.}\ \bibnamefont {Parsons}}, \bibinfo {author} {\bibfnamefont {M.}~\bibnamefont {Kan{\'a}sz-Nagy}}, \bibinfo {author} {\bibfnamefont {R.}~\bibnamefont {Schmidt}}, \bibinfo {author} {\bibfnamefont {F.}~\bibnamefont {Grusdt}}, \bibinfo {author} {\bibfnamefont {E.}~\bibnamefont {Demler}}, \bibinfo {author} {\bibfnamefont {D.}~\bibnamefont {Greif}},\ and\ \bibinfo {author} {\bibfnamefont {M.}~\bibnamefont {Greiner}},\ }\href {https://doi.org/10.1038/nature22362} {\bibfield  {journal} {\bibinfo  {journal} {Nature}\ }\textbf {\bibinfo {volume} {545}},\ \bibinfo {pages} {462} (\bibinfo {year} {2017})}\BibitemShut {NoStop}%
\bibitem [{\citenamefont {Redfield}(1957)}]{Redfield1957}%
  \BibitemOpen
  \bibfield  {author} {\bibinfo {author} {\bibfnamefont {A.~G.}\ \bibnamefont {Redfield}},\ }\href {https://api.semanticscholar.org/CorpusID:33976678} {\bibfield  {journal} {\bibinfo  {journal} {IBM J. Res. Dev.}\ }\textbf {\bibinfo {volume} {1}},\ \bibinfo {pages} {19} (\bibinfo {year} {1957})}\BibitemShut {NoStop}%
\bibitem [{\citenamefont {Tindall}(2021)}]{Tindallthesis}%
  \BibitemOpen
  \bibfield  {author} {\bibinfo {author} {\bibfnamefont {J.}~\bibnamefont {Tindall}},\ }\href@noop {} {\emph {\bibinfo {title} {Realising Complex Quantum States of Matter via Symmetries and Heating (PhD thesis, University of Oxford)}}}\ (\bibinfo {year} {2021})\BibitemShut {NoStop}%
\bibitem [{\citenamefont {Klein}\ \emph {et~al.}(2007)\citenamefont {Klein}, \citenamefont {Bruderer}, \citenamefont {Clark},\ and\ \citenamefont {Jaksch}}]{Klein2007}%
  \BibitemOpen
  \bibfield  {author} {\bibinfo {author} {\bibfnamefont {A.}~\bibnamefont {Klein}}, \bibinfo {author} {\bibfnamefont {M.}~\bibnamefont {Bruderer}}, \bibinfo {author} {\bibfnamefont {S.~R.}\ \bibnamefont {Clark}},\ and\ \bibinfo {author} {\bibfnamefont {D.}~\bibnamefont {Jaksch}},\ }\href {https://doi.org/10.1088/1367-2630/9/11/411} {\bibfield  {journal} {\bibinfo  {journal} {New Journal of Physics}\ }\textbf {\bibinfo {volume} {9}},\ \bibinfo {pages} {411} (\bibinfo {year} {2007})}\BibitemShut {NoStop}%
\bibitem [{\citenamefont {Damanet}\ \emph {et~al.}(2019)\citenamefont {Damanet}, \citenamefont {Daley},\ and\ \citenamefont {Keeling}}]{Damanet19}%
  \BibitemOpen
  \bibfield  {author} {\bibinfo {author} {\bibfnamefont {F.}~\bibnamefont {Damanet}}, \bibinfo {author} {\bibfnamefont {A.~J.}\ \bibnamefont {Daley}},\ and\ \bibinfo {author} {\bibfnamefont {J.}~\bibnamefont {Keeling}},\ }\href {https://doi.org/10.1103/PhysRevA.99.033845} {\bibfield  {journal} {\bibinfo  {journal} {Phys. Rev. A}\ }\textbf {\bibinfo {volume} {99}},\ \bibinfo {pages} {033845} (\bibinfo {year} {2019})}\BibitemShut {NoStop}%
\bibitem [{\citenamefont {Palacino}\ and\ \citenamefont {Keeling}(2021)}]{Palacino21}%
  \BibitemOpen
  \bibfield  {author} {\bibinfo {author} {\bibfnamefont {R.}~\bibnamefont {Palacino}}\ and\ \bibinfo {author} {\bibfnamefont {J.}~\bibnamefont {Keeling}},\ }\href {https://doi.org/10.1103/PhysRevResearch.3.L032016} {\bibfield  {journal} {\bibinfo  {journal} {Phys. Rev. Res.}\ }\textbf {\bibinfo {volume} {3}},\ \bibinfo {pages} {L032016} (\bibinfo {year} {2021})}\BibitemShut {NoStop}%
\bibitem [{\citenamefont {Debecker}\ \emph {et~al.}(2024{\natexlab{a}})\citenamefont {Debecker}, \citenamefont {Martin},\ and\ \citenamefont {Damanet}}]{Debecker23PRL}%
  \BibitemOpen
  \bibfield  {author} {\bibinfo {author} {\bibfnamefont {B.}~\bibnamefont {Debecker}}, \bibinfo {author} {\bibfnamefont {J.}~\bibnamefont {Martin}},\ and\ \bibinfo {author} {\bibfnamefont {F.}~\bibnamefont {Damanet}},\ }\href {https://doi.org/10.1103/PhysRevLett.133.140403} {\bibfield  {journal} {\bibinfo  {journal} {Phys. Rev. Lett.}\ }\textbf {\bibinfo {volume} {133}},\ \bibinfo {pages} {140403} (\bibinfo {year} {2024}{\natexlab{a}})}\BibitemShut {NoStop}%
\bibitem [{\citenamefont {Debecker}\ \emph {et~al.}(2024{\natexlab{b}})\citenamefont {Debecker}, \citenamefont {Martin},\ and\ \citenamefont {Damanet}}]{Debecker23PRA}%
  \BibitemOpen
  \bibfield  {author} {\bibinfo {author} {\bibfnamefont {B.}~\bibnamefont {Debecker}}, \bibinfo {author} {\bibfnamefont {J.}~\bibnamefont {Martin}},\ and\ \bibinfo {author} {\bibfnamefont {F.}~\bibnamefont {Damanet}},\ }\href {https://doi.org/10.1103/PhysRevA.110.042201} {\bibfield  {journal} {\bibinfo  {journal} {Phys. Rev. A}\ }\textbf {\bibinfo {volume} {110}},\ \bibinfo {pages} {042201} (\bibinfo {year} {2024}{\natexlab{b}})}\BibitemShut {NoStop}%
\bibitem [{\citenamefont {Debecker}\ \emph {et~al.}(2025)\citenamefont {Debecker}, \citenamefont {Pausch}, \citenamefont {Louvet}, \citenamefont {Bastin}, \citenamefont {Martin},\ and\ \citenamefont {Damanet}}]{Debecker2025}%
  \BibitemOpen
  \bibfield  {author} {\bibinfo {author} {\bibfnamefont {B.}~\bibnamefont {Debecker}}, \bibinfo {author} {\bibfnamefont {L.}~\bibnamefont {Pausch}}, \bibinfo {author} {\bibfnamefont {J.}~\bibnamefont {Louvet}}, \bibinfo {author} {\bibfnamefont {T.}~\bibnamefont {Bastin}}, \bibinfo {author} {\bibfnamefont {J.}~\bibnamefont {Martin}},\ and\ \bibinfo {author} {\bibfnamefont {F.}~\bibnamefont {Damanet}},\ }\href {https://arxiv.org/abs/2504.11317} {\bibinfo {title} {The role of non-markovian dissipation in quantum phase transitions: tricriticality, spin squeezing, and directional symmetry breaking}} (\bibinfo {year} {2025}),\ \Eprint {https://arxiv.org/abs/2504.11317} {arXiv:2504.11317 [quant-ph]} \BibitemShut {NoStop}%
\bibitem [{\citenamefont {Johansson}\ \emph {et~al.}(2012)\citenamefont {Johansson}, \citenamefont {Nation},\ and\ \citenamefont {Nori}}]{qutip}%
  \BibitemOpen
  \bibfield  {author} {\bibinfo {author} {\bibfnamefont {J.}~\bibnamefont {Johansson}}, \bibinfo {author} {\bibfnamefont {P.}~\bibnamefont {Nation}},\ and\ \bibinfo {author} {\bibfnamefont {F.}~\bibnamefont {Nori}},\ }\bibfield  {journal} {\bibinfo  {journal} {Computer Physics Communications}\ }\href {https://doi.org/10.1016/j.cpc.2012.02.021} {10.1016/j.cpc.2012.02.021} (\bibinfo {year} {2012})\BibitemShut {NoStop}%
\bibitem [{\citenamefont {Bu\v{c}a}\ \emph {et~al.}(2019)\citenamefont {Bu\v{c}a}, \citenamefont {Tindall},\ and\ \citenamefont {Jaksch}}]{Buca2019}%
  \BibitemOpen
  \bibfield  {author} {\bibinfo {author} {\bibfnamefont {B.}~\bibnamefont {Bu\v{c}a}}, \bibinfo {author} {\bibfnamefont {J.}~\bibnamefont {Tindall}},\ and\ \bibinfo {author} {\bibfnamefont {D.}~\bibnamefont {Jaksch}},\ }\href {https://doi.org/10.1038/s41467-019-09757-y} {\bibfield  {journal} {\bibinfo  {journal} {Nature Communications}\ }\textbf {\bibinfo {volume} {10}},\ \bibinfo {pages} {1730} (\bibinfo {year} {2019})}\BibitemShut {NoStop}%
\bibitem [{\citenamefont {Lucchesi}\ and\ \citenamefont {Chiofalo}(2019)}]{Lucchesi2019}%
  \BibitemOpen
  \bibfield  {author} {\bibinfo {author} {\bibfnamefont {L.}~\bibnamefont {Lucchesi}}\ and\ \bibinfo {author} {\bibfnamefont {M.~L.}\ \bibnamefont {Chiofalo}},\ }\href {https://doi.org/10.1103/PhysRevLett.123.060406} {\bibfield  {journal} {\bibinfo  {journal} {Phys. Rev. Lett.}\ }\textbf {\bibinfo {volume} {123}},\ \bibinfo {pages} {060406} (\bibinfo {year} {2019})}\BibitemShut {NoStop}%
\bibitem [{\citenamefont {Kruchinin}\ \emph {et~al.}(2010)\citenamefont {Kruchinin}, \citenamefont {Nagao},\ and\ \citenamefont {Aono}}]{Kruchinin2010}%
  \BibitemOpen
  \bibfield  {author} {\bibinfo {author} {\bibfnamefont {S.}~\bibnamefont {Kruchinin}}, \bibinfo {author} {\bibfnamefont {H.}~\bibnamefont {Nagao}},\ and\ \bibinfo {author} {\bibfnamefont {S.}~\bibnamefont {Aono}},\ }\href@noop {} {\emph {\bibinfo {title} {Modern Aspects of Superconductivity}}}\ (\bibinfo  {publisher} {WORLD SCIENTIFIC},\ \bibinfo {year} {2010})\BibitemShut {NoStop}%
\bibitem [{\citenamefont {Daley}\ \emph {et~al.}(2004)\citenamefont {Daley}, \citenamefont {Fedichev},\ and\ \citenamefont {Zoller}}]{Daley2004}%
  \BibitemOpen
  \bibfield  {author} {\bibinfo {author} {\bibfnamefont {A.~J.}\ \bibnamefont {Daley}}, \bibinfo {author} {\bibfnamefont {P.~O.}\ \bibnamefont {Fedichev}},\ and\ \bibinfo {author} {\bibfnamefont {P.}~\bibnamefont {Zoller}},\ }\href {https://doi.org/10.1103/PhysRevA.69.022306} {\bibfield  {journal} {\bibinfo  {journal} {Phys. Rev. A}\ }\textbf {\bibinfo {volume} {69}},\ \bibinfo {pages} {022306} (\bibinfo {year} {2004})}\BibitemShut {NoStop}%
\bibitem [{\citenamefont {Bruderer}\ \emph {et~al.}(2007)\citenamefont {Bruderer}, \citenamefont {Klein}, \citenamefont {Clark},\ and\ \citenamefont {Jaksch}}]{Bruderer2007}%
  \BibitemOpen
  \bibfield  {author} {\bibinfo {author} {\bibfnamefont {M.}~\bibnamefont {Bruderer}}, \bibinfo {author} {\bibfnamefont {A.}~\bibnamefont {Klein}}, \bibinfo {author} {\bibfnamefont {S.~R.}\ \bibnamefont {Clark}},\ and\ \bibinfo {author} {\bibfnamefont {D.}~\bibnamefont {Jaksch}},\ }\href {https://doi.org/10.1103/PhysRevA.76.011605} {\bibfield  {journal} {\bibinfo  {journal} {Phys. Rev. A}\ }\textbf {\bibinfo {volume} {76}},\ \bibinfo {pages} {011605} (\bibinfo {year} {2007})}\BibitemShut {NoStop}%
\bibitem [{\citenamefont {Chin}\ \emph {et~al.}(2010)\citenamefont {Chin}, \citenamefont {Grimm}, \citenamefont {Julienne},\ and\ \citenamefont {Tiesinga}}]{Chin2010}%
  \BibitemOpen
  \bibfield  {author} {\bibinfo {author} {\bibfnamefont {C.}~\bibnamefont {Chin}}, \bibinfo {author} {\bibfnamefont {R.}~\bibnamefont {Grimm}}, \bibinfo {author} {\bibfnamefont {P.}~\bibnamefont {Julienne}},\ and\ \bibinfo {author} {\bibfnamefont {E.}~\bibnamefont {Tiesinga}},\ }\href {https://doi.org/10.1103/RevModPhys.82.1225} {\bibfield  {journal} {\bibinfo  {journal} {Rev. Mod. Phys.}\ }\textbf {\bibinfo {volume} {82}},\ \bibinfo {pages} {1225} (\bibinfo {year} {2010})}\BibitemShut {NoStop}%
\bibitem [{\citenamefont {Mandel}\ \emph {et~al.}(2003)\citenamefont {Mandel}, \citenamefont {Greiner}, \citenamefont {Widera}, \citenamefont {Rom}, \citenamefont {H{\"a}nsch},\ and\ \citenamefont {Bloch}}]{Mandel2003}%
  \BibitemOpen
  \bibfield  {author} {\bibinfo {author} {\bibfnamefont {O.}~\bibnamefont {Mandel}}, \bibinfo {author} {\bibfnamefont {M.}~\bibnamefont {Greiner}}, \bibinfo {author} {\bibfnamefont {A.}~\bibnamefont {Widera}}, \bibinfo {author} {\bibfnamefont {T.}~\bibnamefont {Rom}}, \bibinfo {author} {\bibfnamefont {T.~W.}\ \bibnamefont {H{\"a}nsch}},\ and\ \bibinfo {author} {\bibfnamefont {I.}~\bibnamefont {Bloch}},\ }\href {https://api.semanticscholar.org/CorpusID:37487570} {\bibfield  {journal} {\bibinfo  {journal} {Physical review letters}\ }\textbf {\bibinfo {volume} {91 1}},\ \bibinfo {pages} {010407} (\bibinfo {year} {2003})}\BibitemShut {NoStop}%
\bibitem [{\citenamefont {Yang}\ \emph {et~al.}(2017)\citenamefont {Yang}, \citenamefont {Dai}, \citenamefont {Sun}, \citenamefont {Reingruber}, \citenamefont {Yuan},\ and\ \citenamefont {Pan}}]{Yang2019}%
  \BibitemOpen
  \bibfield  {author} {\bibinfo {author} {\bibfnamefont {B.}~\bibnamefont {Yang}}, \bibinfo {author} {\bibfnamefont {H.-N.}\ \bibnamefont {Dai}}, \bibinfo {author} {\bibfnamefont {H.}~\bibnamefont {Sun}}, \bibinfo {author} {\bibfnamefont {A.}~\bibnamefont {Reingruber}}, \bibinfo {author} {\bibfnamefont {Z.-S.}\ \bibnamefont {Yuan}},\ and\ \bibinfo {author} {\bibfnamefont {J.-W.}\ \bibnamefont {Pan}},\ }\href {https://doi.org/10.1103/PhysRevA.96.011602} {\bibfield  {journal} {\bibinfo  {journal} {Phys. Rev. A}\ }\textbf {\bibinfo {volume} {96}},\ \bibinfo {pages} {011602} (\bibinfo {year} {2017})}\BibitemShut {NoStop}%
\bibitem [{\citenamefont {Bernier}\ \emph {et~al.}(2013)\citenamefont {Bernier}, \citenamefont {Barmettler}, \citenamefont {Poletti},\ and\ \citenamefont {Kollath}}]{Bernier2013}%
  \BibitemOpen
  \bibfield  {author} {\bibinfo {author} {\bibfnamefont {J.-S.}\ \bibnamefont {Bernier}}, \bibinfo {author} {\bibfnamefont {P.}~\bibnamefont {Barmettler}}, \bibinfo {author} {\bibfnamefont {D.}~\bibnamefont {Poletti}},\ and\ \bibinfo {author} {\bibfnamefont {C.}~\bibnamefont {Kollath}},\ }\href {https://doi.org/10.1103/PhysRevA.87.063608} {\bibfield  {journal} {\bibinfo  {journal} {Phys. Rev. A}\ }\textbf {\bibinfo {volume} {87}},\ \bibinfo {pages} {063608} (\bibinfo {year} {2013})}\BibitemShut {NoStop}%
\bibitem [{\citenamefont {Shin}\ \emph {et~al.}(2006)\citenamefont {Shin}, \citenamefont {Zwierlein}, \citenamefont {Schunck}, \citenamefont {Schirotzek},\ and\ \citenamefont {Ketterle}}]{Shin2006}%
  \BibitemOpen
  \bibfield  {author} {\bibinfo {author} {\bibfnamefont {Y.}~\bibnamefont {Shin}}, \bibinfo {author} {\bibfnamefont {M.~W.}\ \bibnamefont {Zwierlein}}, \bibinfo {author} {\bibfnamefont {C.~H.}\ \bibnamefont {Schunck}}, \bibinfo {author} {\bibfnamefont {A.}~\bibnamefont {Schirotzek}},\ and\ \bibinfo {author} {\bibfnamefont {W.}~\bibnamefont {Ketterle}},\ }\href {https://doi.org/10.1103/PhysRevLett.97.030401} {\bibfield  {journal} {\bibinfo  {journal} {Phys. Rev. Lett.}\ }\textbf {\bibinfo {volume} {97}},\ \bibinfo {pages} {030401} (\bibinfo {year} {2006})}\BibitemShut {NoStop}%
\end{thebibliography}%

\end{document}